\documentclass[referee]{aa} 
\usepackage{graphicx}
\graphicspath{{figures/}}
\usepackage{amsmath}
\usepackage{txfonts}
\usepackage{fourier}
\usepackage{url}
\usepackage{longtable}
\usepackage{tikz}
\usepackage{chemformula}
\usepackage{placeins}

\usepackage{natbib}
\usepackage{acronym}
\acrodef{PC}[PC]{Phase Curve}
\acrodef{LC}[LC]{Light Curve}
\acrodef{LD}[LD]{Limb Darkening}
\acrodef{GP}[GP]{Gaussian Process}
\acrodef{MCMC}[MCMC]{Markov Chain Monte Carlo}
\acrodef{HJ}[HJ]{Hot Jupiter}
\acrodef{FFI}[FFI]{Full Frame Images}
\acrodef{TPF}[TPF]{Target Pixel File}
\acrodef{CBV}[CBV]{Co-trending Basis Vectors}
\acrodef{DRP}[DRP]{Data Reduction Pipeline}
\acrodef{IRFM}[IRFM]{Infra-Red Flux Method}
\acrodef{SED}[SED]{Spectral Energy Distribution}
\acrodef{FAP}[FAP]{False Alarm Probability}
\acrodef{MAP}[MAP]{Maximum A Posteriori}
\acrodef{GLS}[GLS]{Generalized Lomb-Scargle}
\acrodef{PSF}[PSF]{Point Spread Function}
 
\begin{document}

\newcommand{\teff}{T$_{\rm eff}$}
\newcommand{\logg}{$\log${(g)}}
\newcommand{\wcsob}{WASP-178~b}
\newcommand{\wcso}{WASP-178}
\newcommand{\cheops}{CHEOPS}
\newcommand{\tess}{TESS}
\newcommand{\uvis}{UVIS}
\newcommand{\aas}{AAS}
\newcommand{\vsini}{\ensuremath{v \sin i_\star}\xspace}
\newcommand{\ageo}{$A_{\rm g}$}
\newcommand{\asph}{$A_{\rm S}$}
\newcommand{\abond}{$A_{\rm B}$}

\let\orgautoref\autoref
\makeatletter
\renewcommand*\aa@pageof{, page \thepage{} of \pageref*{LastPage}}
\def\instrefs#1{{\def\scsep{\def\scsep{,}}\@for\w:=#1\do{\scsep\ref{inst:\w}}}}
\makeatother

   \title{Constraining the reflective properties of \wcsob\ using \cheops\ photometry.\thanks{The \cheops\ program ID is CH\_PR100016.}\thanks{\textbf{The CHEOPS photometric data used in this work are only available in electronic form
at the CDS via anonymous ftp to cdsarc.cds.unistra.fr (130.79.128.5)
or via https://cdsarc.cds.unistra.fr/cgi-bin/qcat?J/A+A/}}}


\author{
I. Pagano\inst{1} $^{\href{https://orcid.org/0000-0001-9573-4928}{\includegraphics[scale=0.5]{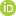}}}$, 
G. Scandariato\inst{1} $^{\href{https://orcid.org/0000-0003-2029-0626}{\includegraphics[scale=0.5]{figures/orcid.jpg}}}$, 
V. Singh\inst{1} $^{\href{https://orcid.org/0000-0002-7485-6309}{\includegraphics[scale=0.5]{figures/orcid.jpg}}}$, 
M. Lendl\inst{2} $^{\href{https://orcid.org/0000-0001-9699-1459}{\includegraphics[scale=0.5]{figures/orcid.jpg}}}$, 
D. Queloz\inst{3,4} $^{\href{https://orcid.org/0000-0002-3012-0316}{\includegraphics[scale=0.5]{figures/orcid.jpg}}}$, 
A. E. Simon\inst{5} $^{\href{https://orcid.org/0000-0001-9773-2600}{\includegraphics[scale=0.5]{figures/orcid.jpg}}}$, 
S. G. Sousa\inst{6} $^{\href{https://orcid.org/0000-0001-9047-2965}{\includegraphics[scale=0.5]{figures/orcid.jpg}}}$, 
A. Brandeker\inst{7} $^{\href{https://orcid.org/0000-0002-7201-7536}{\includegraphics[scale=0.5]{figures/orcid.jpg}}}$, 
A. Collier Cameron\inst{8} $^{\href{https://orcid.org/0000-0002-8863-7828}{\includegraphics[scale=0.5]{figures/orcid.jpg}}}$, 
S. Sulis\inst{9} $^{\href{https://orcid.org/0000-0001-8783-526X}{\includegraphics[scale=0.5]{figures/orcid.jpg}}}$, 
V. Van Grootel\inst{10} $^{\href{https://orcid.org/0000-0003-2144-4316}{\includegraphics[scale=0.5]{figures/orcid.jpg}}}$, 
T. G. Wilson\inst{8} $^{\href{https://orcid.org/0000-0001-8749-1962}{\includegraphics[scale=0.5]{figures/orcid.jpg}}}$, 
Y. Alibert\inst{11,5} $^{\href{https://orcid.org/0000-0002-4644-8818}{\includegraphics[scale=0.5]{figures/orcid.jpg}}}$, 
R. Alonso\inst{12,13} $^{\href{https://orcid.org/0000-0001-8462-8126}{\includegraphics[scale=0.5]{figures/orcid.jpg}}}$, 
G. Anglada\inst{14,15} $^{\href{https://orcid.org/0000-0002-3645-5977}{\includegraphics[scale=0.5]{figures/orcid.jpg}}}$, 
T. Bárczy\inst{16} $^{\href{https://orcid.org/0000-0002-7822-4413}{\includegraphics[scale=0.5]{figures/orcid.jpg}}}$, 
D. Barrado Navascues\inst{17} $^{\href{https://orcid.org/0000-0002-5971-9242}{\includegraphics[scale=0.5]{figures/orcid.jpg}}}$, 
S. C. C. Barros\inst{6,18} $^{\href{https://orcid.org/0000-0003-2434-3625}{\includegraphics[scale=0.5]{figures/orcid.jpg}}}$, 
W. Baumjohann\inst{19} $^{\href{https://orcid.org/0000-0001-6271-0110}{\includegraphics[scale=0.5]{figures/orcid.jpg}}}$, 
M. Beck\inst{2} $^{\href{https://orcid.org/0000-0003-3926-0275}{\includegraphics[scale=0.5]{figures/orcid.jpg}}}$, 
T. Beck\inst{5}, 
W. Benz\inst{5,11} $^{\href{https://orcid.org/0000-0001-7896-6479}{\includegraphics[scale=0.5]{figures/orcid.jpg}}}$, 
N. Billot\inst{2} $^{\href{https://orcid.org/0000-0003-3429-3836}{\includegraphics[scale=0.5]{figures/orcid.jpg}}}$, 
X. Bonfils\inst{20} $^{\href{https://orcid.org/0000-0001-9003-8894}{\includegraphics[scale=0.5]{figures/orcid.jpg}}}$, 
L. Borsato\inst{21} $^{\href{https://orcid.org/0000-0003-0066-9268}{\includegraphics[scale=0.5]{figures/orcid.jpg}}}$, 
C. Broeg\inst{5,11} $^{\href{https://orcid.org/0000-0001-5132-2614}{\includegraphics[scale=0.5]{figures/orcid.jpg}}}$, 
G. Bruno\inst{1} $^{\href{https://orcid.org/0000-0002-3288-0802}{\includegraphics[scale=0.5]{figures/orcid.jpg}}}$, 
L. Carone\inst{19}, 
S. Charnoz\inst{22} $^{\href{https://orcid.org/0000-0002-7442-491X}{\includegraphics[scale=0.5]{figures/orcid.jpg}}}$, 
C. Corral van Damme\inst{23}, 
Sz. Csizmadia\inst{24} $^{\href{https://orcid.org/0000-0001-6803-9698}{\includegraphics[scale=0.5]{figures/orcid.jpg}}}$, 
P. E. Cubillos\inst{25,19}, 
M. B. Davies\inst{26} $^{\href{https://orcid.org/0000-0001-6080-1190}{\includegraphics[scale=0.5]{figures/orcid.jpg}}}$, 
M. Deleuil\inst{9} $^{\href{https://orcid.org/0000-0001-6036-0225}{\includegraphics[scale=0.5]{figures/orcid.jpg}}}$, 
A. Deline\inst{2}, 
L. Delrez\inst{27,10} $^{\href{https://orcid.org/0000-0001-6108-4808}{\includegraphics[scale=0.5]{figures/orcid.jpg}}}$, 
O. D. S. Demangeon\inst{6,18} $^{\href{https://orcid.org/0000-0001-7918-0355}{\includegraphics[scale=0.5]{figures/orcid.jpg}}}$, 
B.-O. Demory\inst{11,5} $^{\href{https://orcid.org/0000-0002-9355-5165}{\includegraphics[scale=0.5]{figures/orcid.jpg}}}$, 
D. Ehrenreich\inst{2,28} $^{\href{https://orcid.org/0000-0001-9704-5405}{\includegraphics[scale=0.5]{figures/orcid.jpg}}}$, 
A. Erikson\inst{24}, 
A. Fortier\inst{5,11} $^{\href{https://orcid.org/0000-0001-8450-3374}{\includegraphics[scale=0.5]{figures/orcid.jpg}}}$, 
L. Fossati\inst{19} $^{\href{https://orcid.org/0000-0003-4426-9530}{\includegraphics[scale=0.5]{figures/orcid.jpg}}}$, 
M. Fridlund\inst{29,30} $^{\href{https://orcid.org/0000-0002-0855-8426}{\includegraphics[scale=0.5]{figures/orcid.jpg}}}$, 
D. Gandolfi\inst{31} $^{\href{https://orcid.org/0000-0001-8627-9628}{\includegraphics[scale=0.5]{figures/orcid.jpg}}}$, 
M. Gillon\inst{27} $^{\href{https://orcid.org/0000-0003-1462-7739}{\includegraphics[scale=0.5]{figures/orcid.jpg}}}$, 
M. Güdel\inst{32}, 
M. N. Günther\inst{23}, 
Ch. Helling\inst{19,51}, 
S. Hoyer\inst{9} $^{\href{https://orcid.org/0000-0003-3477-2466}{\includegraphics[scale=0.5]{figures/orcid.jpg}}}$, 
K. G. Isaak\inst{33} $^{\href{https://orcid.org/0000-0001-8585-1717}{\includegraphics[scale=0.5]{figures/orcid.jpg}}}$, 
L. L. Kiss\inst{34,35}, 
E. Kopp\inst{36}, 
K. W. F. Lam\inst{24} $^{\href{https://orcid.org/0000-0002-9910-6088}{\includegraphics[scale=0.5]{figures/orcid.jpg}}}$, 
J. Laskar\inst{37} $^{\href{https://orcid.org/0000-0003-2634-789X}{\includegraphics[scale=0.5]{figures/orcid.jpg}}}$, 
A. Lecavelier des Etangs\inst{38} $^{\href{https://orcid.org/0000-0002-5637-5253}{\includegraphics[scale=0.5]{figures/orcid.jpg}}}$, 
D. Magrin\inst{21} $^{\href{https://orcid.org/0000-0003-0312-313X}{\includegraphics[scale=0.5]{figures/orcid.jpg}}}$, 
P. F. L. Maxted\inst{39} $^{\href{https://orcid.org/0000-0003-3794-1317}{\includegraphics[scale=0.5]{figures/orcid.jpg}}}$, 
C. Mordasini\inst{5,11}, 
M. Munari\inst{1} $^{\href{https://orcid.org/0000-0003-0990-050X}{\includegraphics[scale=0.5]{figures/orcid.jpg}}}$, 
V. Nascimbeni\inst{21} $^{\href{https://orcid.org/0000-0001-9770-1214}{\includegraphics[scale=0.5]{figures/orcid.jpg}}}$, 
G. Olofsson\inst{7} $^{\href{https://orcid.org/0000-0003-3747-7120}{\includegraphics[scale=0.5]{figures/orcid.jpg}}}$, 
R. Ottensamer\inst{32}, 
E. Pallé\inst{12,13} $^{\href{https://orcid.org/0000-0003-0987-1593}{\includegraphics[scale=0.5]{figures/orcid.jpg}}}$, 
G. Peter\inst{40} $^{\href{https://orcid.org/0000-0001-6101-2513}{\includegraphics[scale=0.5]{figures/orcid.jpg}}}$, 
G. Piotto\inst{21,41} $^{\href{https://orcid.org/0000-0002-9937-6387}{\includegraphics[scale=0.5]{figures/orcid.jpg}}}$, 
D. Pollacco\inst{42}, 
R. Ragazzoni\inst{21,41} $^{\href{https://orcid.org/0000-0002-7697-5555}{\includegraphics[scale=0.5]{figures/orcid.jpg}}}$, 
N. Rando\inst{23}, 
H. Rauer\inst{24,43,44} $^{\href{https://orcid.org/0000-0002-6510-1828}{\includegraphics[scale=0.5]{figures/orcid.jpg}}}$, 
C. Reimers\inst{45}, 
I. Ribas\inst{14,15} $^{\href{https://orcid.org/0000-0002-6689-0312}{\includegraphics[scale=0.5]{figures/orcid.jpg}}}$, 
M. Rieder\inst{46,11}, 
N. C. Santos\inst{6,18} $^{\href{https://orcid.org/0000-0003-4422-2919}{\includegraphics[scale=0.5]{figures/orcid.jpg}}}$, 
D. Ségransan\inst{2} $^{\href{https://orcid.org/0000-0003-2355-8034}{\includegraphics[scale=0.5]{figures/orcid.jpg}}}$, 
A. M. S. Smith\inst{24} $^{\href{https://orcid.org/0000-0002-2386-4341}{\includegraphics[scale=0.5]{figures/orcid.jpg}}}$, 
M. Stalport\inst{47}, 
M. Steller\inst{19} $^{\href{https://orcid.org/0000-0003-2459-6155}{\includegraphics[scale=0.5]{figures/orcid.jpg}}}$, 
Gy. M. Szabó\inst{48,49}, 
N. Thomas\inst{5}, 
S. Udry\inst{2} $^{\href{https://orcid.org/0000-0001-7576-6236}{\includegraphics[scale=0.5]{figures/orcid.jpg}}}$, 
J. Venturini\inst{2} $^{\href{https://orcid.org/0000-0001-9527-2903}{\includegraphics[scale=0.5]{figures/orcid.jpg}}}$, 
N. A. Walton\inst{50} $^{\href{https://orcid.org/0000-0003-3983-8778}{\includegraphics[scale=0.5]{figures/orcid.jpg}}}$
}

\institute{
\label{inst:1} INAF, Osservatorio Astrofisico di Catania, Via S. Sofia 78, 95123 Catania, Italy \and
\label{inst:2} Observatoire Astronomique de l'Université de Genève, Chemin Pegasi 51, CH-1290 Versoix, Switzerland \and
\label{inst:3} ETH Zurich, Department of Physics, Wolfgang-Pauli-Strasse 2, CH-8093 Zurich, Switzerland \and
\label{inst:4} Cavendish Laboratory, JJ Thomson Avenue, Cambridge CB3 0HE, UK \and
\label{inst:5} Physikalisches Institut, University of Bern, Gesellschaftsstrasse 6, 3012 Bern, Switzerland \and
\label{inst:6} Instituto de Astrofisica e Ciencias do Espaco, Universidade do Porto, CAUP, Rua das Estrelas, 4150-762 Porto, Portugal \and
\label{inst:7} Department of Astronomy, Stockholm University, AlbaNova University Center, 10691 Stockholm, Sweden \and
\label{inst:8} Centre for Exoplanet Science, SUPA School of Physics and Astronomy, University of St Andrews, North Haugh, St Andrews KY16 9SS, UK \and
\label{inst:9} Aix Marseille Univ, CNRS, CNES, LAM, 38 rue Frédéric Joliot-Curie, 13388 Marseille, France \and
\label{inst:10} Space sciences, Technologies and Astrophysics Research (STAR) Institute, Université de Liège, Allée du 6 Août 19C, 4000 Liège, Belgium \and
\label{inst:11} Center for Space and Habitability, University of Bern, Gesellschaftsstrasse 6, 3012 Bern, Switzerland \and
\label{inst:12} Instituto de Astrofisica de Canarias, Via Lactea s/n, 38200 La Laguna, Tenerife, Spain \and
\label{inst:13} Departamento de Astrofisica, Universidad de La Laguna, Astrofísico Francisco Sanchez s/n, 38206 La Laguna, Tenerife, Spain \and
\label{inst:14} Institut de Ciencies de l'Espai (ICE, CSIC), Campus UAB, Can Magrans s/n, 08193 Bellaterra, Spain \and
\label{inst:15} Institut d'Estudis Espacials de Catalunya (IEEC), Gran Capità 2-4, 08034 Barcelona, Spain \and
\label{inst:16} Admatis, 5. Kandó Kálmán Street, 3534 Miskolc, Hungary \and
\label{inst:17} Depto. de Astrofisica, Centro de Astrobiologia (CSIC-INTA), ESAC campus, 28692 Villanueva de la Cañada (Madrid), Spain \and
\label{inst:18} Departamento de Fisica e Astronomia, Faculdade de Ciencias, Universidade do Porto, Rua do Campo Alegre, 4169-007 Porto, Portugal \and
\label{inst:19} Space Research Institute, Austrian Academy of Sciences, Schmiedlstrasse 6, A-8042 Graz, Austria \and
\label{inst:20} Université Grenoble Alpes, CNRS, IPAG, 38000 Grenoble, France \and
\label{inst:21} INAF, Osservatorio Astronomico di Padova, Vicolo dell'Osservatorio 5, 35122 Padova, Italy \and
\label{inst:22} Université de Paris Cité, Institut de physique du globe de Paris, CNRS, 1 Rue Jussieu, F-75005 Paris, France \and
\label{inst:23} ESTEC, European Space Agency, Keplerlaan 1, 2201AZ, Noordwijk, NL \and
\label{inst:24} Institute of Planetary Research, German Aerospace Center (DLR), Rutherfordstrasse 2, 12489 Berlin, Germany \and
\label{inst:25} INAF, Osservatorio Astrofisico di Torino, Via Osservatorio, 20, I-10025 Pino Torinese To, Italy \and
\label{inst:26} Centre for Mathematical Sciences, Lund University, Box 118, 221 00 Lund, Sweden \and
\label{inst:27} Astrobiology Research Unit, Université de Liège, Allée du 6 Août 19C, B-4000 Liège, Belgium \and
\label{inst:28} Centre Vie dans l'Univers, Facult\'e des sciences, Universit\'e de Gen\`eve, Quai Ernest-Ansermet 30, CH-1211 Gen\`eve 4, Switzerland \and
\label{inst:29} Leiden Observatory, University of Leiden, PO Box 9513, 2300 RA Leiden, The Netherlands \and
\label{inst:30} Department of Space, Earth and Environment, Chalmers University of Technology, Onsala Space Observatory, 439 92 Onsala, Sweden \and
\label{inst:31} Dipartimento di Fisica, Universita degli Studi di Torino, via Pietro Giuria 1, I-10125, Torino, Italy \and
\label{inst:32} Department of Astrophysics, University of Vienna, Türkenschanzstrasse 17, 1180 Vienna, Austria \and
\label{inst:33} Science and Operations Department - Science Division (SCI-SC), Directorate of Science, European Space Agency (ESA), European Space Research and Technology Centre (ESTEC), Keplerlaan 1, 2201-AZ Noordwijk, The Netherlands \and
\label{inst:34} Konkoly Observatory, Research Centre for Astronomy and Earth Sciences, 1121 Budapest, Konkoly Thege Miklós út 15-17, Hungary \and
\label{inst:35} ELTE E\"otv\"os Lor\'and University, Institute of Physics, P\'azm\'any P\'eter s\'et\'any 1/A, 1117 Budapest, Hungary \and
\label{inst:36} German Aerospace Center (DLR), Institute of Optical Sensor Systems, Rutherfordstraße 2, 12489 Berlin \and
\label{inst:37} IMCCE, UMR8028 CNRS, Observatoire de Paris, PSL Univ., Sorbonne Univ., 77 av. Denfert-Rochereau, 75014 Paris, France \and
\label{inst:38} Institut d'astrophysique de Paris, UMR7095 CNRS, Université Pierre \& Marie Curie, 98bis blvd. Arago, 75014 Paris, France \and
\label{inst:39} Astrophysics Group, Lennard Jones Building, Keele University, Staffordshire, ST5 5BG, United Kingdom \and
\label{inst:40} Institute of Optical Sensor Systems, German Aerospace Center (DLR), Rutherfordstrasse 2, 12489 Berlin, Germany \and
\label{inst:41} Dipartimento di Fisica e Astronomia "Galileo Galilei", Universita degli Studi di Padova, Vicolo dell'Osservatorio 3, 35122 Padova, Italy \and
\label{inst:42} Department of Physics, University of Warwick, Gibbet Hill Road, Coventry CV4 7AL, United Kingdom \and
\label{inst:43} Zentrum für Astronomie und Astrophysik, Technische Universität Berlin, Hardenbergstr. 36, D-10623 Berlin, Germany \and
\label{inst:44} Institut fuer Geologische Wissenschaften, Freie Universitaet Berlin, Maltheserstrasse 74-100,12249 Berlin, Germany \and
\label{inst:45} Department of Astrophysics, University of Vienna, Tuerkenschanzstrasse 17, 1180 Vienna, Austria \and
\label{inst:46} Physikalisches Institut, University of Bern, Sidlerstrasse 5, 3012 Bern, Switzerland \and
\label{inst:47} Université de Liège, Allée du 6 Août 19C, 4000 Liège, Belgium \and
\label{inst:48} ELTE E\"otv\"os Lor\'and University, Gothard Astrophysical Observatory, 9700 Szombathely, Szent Imre h. u. 112, Hungary \and
\label{inst:49} MTA-ELTE Exoplanet Research Group, 9700 Szombathely, Szent Imre h. u. 112, Hungary \and
\label{inst:50} Institute of Astronomy, University of Cambridge, Madingley Road, Cambridge, CB3 0HA, United Kingdom \and
\label{inst:51} Institute for Theoretical Physics and Computational Physics, Graz University of Technology, Petersgasse 16, 8010 Graz, Austria
}


 
  \abstract{Multiwavelength photometry of the secondary eclipses of extrasolar planets is able to disentangle the reflected and thermally emitted light radiated from the planetary dayside. This leads to the measurement of the planetary geometric albedo $A_g$, which is an indicator of the presence of clouds in the atmosphere, and the recirculation efficiency $\epsilon$, which quantifies the energy transport within the atmosphere.}
  {In this work we aim to measure $A_g$ and $\epsilon$ for the planet \wcsob, a highly irradiated giant planet with an estimated equilibrium temperature of 2450~K.}
  {We analyzed archival spectra and the light curves collected by \cheops\ and \tess\ to characterize the host \wcso, refine the ephemeris of the system and measure the eclipse depth in the passbands of the two respective telescopes.}
  {We measured a marginally significant eclipse depth of 70$\pm$40~ppm in the \tess\ passband and statistically significant depth of 70$\pm$20~ppm in the \cheops\ passband.}
  {Combining the eclipse depth measurement in the \cheops\ {($\lambda_{\rm eff}=6300~\AA$)} and \tess\ {($\lambda_{\rm eff}=8000~\AA$)} passbands we constrained the dayside brightness temperature of \wcsob\ in the 2250-2800~K interval. The geometric albedo 0.1<$\rm A_g$<0.35 is in general agreement with the picture of poorly reflective giant planets, while the recirculation efficiency $\epsilon>$0.7 makes \wcsob\ an interesting laboratory to test the current heat recirculation models.}


\authorrunning{Pagano et al.}


   \maketitle
%






\section{Introduction}


In the last few years we have gained access to a detailed characterization of exoplanets. Ground-based and space-born instrumentation have progressed such to allow the analysis of the atmosphere of exoplanets, in terms of thermodynamic state and chemical composition \citep{Sing2016,Giacobbe2021}. In particular, current photometric facilities allow to observe the secondary eclipse of giant exoplanets in close orbits \citep{Stevenson2017,Lendl2020,Wong2020,Singh2022}. In this research area, \cheops\ \citep{Benz2021} is bringing a valuable contribution given its ultra-high photometric accuracy capabilities \citep{Lendl2020,Deline2022,Hooton2021,Brandeker2022,Parviainen2022,Scandariato2022,Demory2023}.

The depth of the eclipse quantifies the brightness of the planetary dayside with respect to its parent star. Depending on the temperature of the planet and the photometric band used for the observations, the eclipse depth provides insight into the reflectivity and energy redistribution of the atmosphere.

\wcsob\ (HD~134004~b) is a hot Jupiter discovered by \citet{Hellier2019} and independently announced by \citet{Rodriguez2020} as KELT-26~b. It orbits an A1 IV-V dwarf star (Table~\ref{tab:parameters}) at a distance of 7 stellar radii: these features place \wcsob\ among the planets that receive the highest energy budget from their respective host stars. It is thus an interesting laboratory to test atmospheric models in the presence of extreme irradiation.

In this paper we analyze the eclipse depths measured using the data collected by the \cheops\ and \tess\ space telescopes. It is organized as follows. Sect.~\ref{sec:observations} describes the data acquisition and reduction, while in Sect.~\ref{sec:spectroscopy} we describe how we derive the stellar radius, mass and age. In Sect.~\ref{sec:modeling} we update the orbital solution of \wcsob, put an upper limit on the eclipse depth using \tess\ data and get a significant detection using \cheops\ photometry. Finally, in Sect.~\ref{sec:discussion} we discuss the implication of the extracted eclipse signal in terms of geometric albedo and atmospheric recirculation efficiency.

\begin{table*}
\caption{Stellar and system parameters.}             
\label{tab:parameters}      
\centering          
\begin{tabular}{l c c c r}     
\hline\hline       
Parameter & Symbol & Units & Value & Ref.\\
\hline
V mag & & & 9.95 & \citet{Hellier2019}\\
Spectral Type & & & A1 IV-V & \citet{Hellier2019}\\
Effective temperature & $\rm T_{eff}$ & K & $9350\pm150$ & \citet{Hellier2019}\\
Surface gravity & $\log g$ & --- & $4.35\pm0.15$ & \citet{Hellier2019}\\
Metallicity & [Fe/H] & --- & $0.21\pm0.16$ & \citet{Hellier2019}\\
Projected rotational velocity & $v\sin{i}$ & km/s & $8.2\pm0.6$ & \citet{Hellier2019}\\
Stellar radius & $\rm R_\star$ & $\rm R_\sun$ & $1.722\pm0.020$ & this work\\
Stellar mass & $\rm M_\star$ & $\rm M_\sun$ & $2.169_{-0.089}^{+0.083}$ & this work\\
Stellar age & $t_\star$ & Gyr & $0.05_{-0.05}^{+0.06}$ & this work\\
Radial velocity semi-amplitude & $\rm K_{RV}$ & m/s & $139\pm9$ & \citet{Hellier2019}\\
\hline                  
\end{tabular}
\end{table*}

\section{Observations and data reduction}\label{sec:observations}

\subsection{\textit{TESS} observations}\label{sec:TESSobs}

\tess\ \citep{Ricker2014} observed the \wcso\ system in sector 11 (from 2019 April 23 to 2019 May 20) with a 30 min cadence and in sector 38 (from 2020 April 29 to 2020 May 26) with a 2 min cadence. \citet{Rodriguez2020} claimed a modulation with period of 0.369526~days in the photometry of sector 11, interpreting it as a $\delta$~Scuti pulsation mode. Later, \citet{Lothringer2022} found out that the \tess\ photometry is heavily contaminated by the background eclipsing binary ASASSN-V J150908.07-424253.6, located at a projected distance of 50.4\arcsec\ from the target, whose orbital period matches the periodicity in the photometry of \wcso. In order to optimize the photometric extraction and avoid the background variable contamination, we extracted the \ac{LC} corresponding to each pixel in the aperture mask defined by the \tess\ pipeline, and we excluded the pixels for which the periodogram shows a peak at the same orbital period of the binary system. We thus re-extracted the photometry by retrieving the calibrated \acp{FFI} and \acp{TPF} for sector 11 and 38 respectively. We used a custom extraction pipeline combined with the default quality bitmask. The extracted \acp{LC} were then background-corrected after determining the sky level using custom background masks on the \acp{FFI} and \acp{TPF}. A principal component analysis was then conducted on the pixels in these background masks across all frames in order to measure the flux contribution of scattered light in the \tess\ cameras. We detrended the data using these principal components as a linear model. Lastly, we further corrected for any photometric trends due to spacecraft pointing jitter by retrieving the \acp{CBV} and two-second cadence engineering quaternion measurements for the specific cameras \wcso\ was observed in. For each sector we computed the mean of the quaternions over the length of a science observation, i.e. 30\,min for the \acp{FFI} and 2\,min for the \acp{TPF}. We subsequently used these vectors along with the \acp{CBV} to detrend the \tess\ photometry in a similar manner as was done in \citet{Delrez2021}.

We clip out the photometry taken before BJD 2459334.7 and in the windows 2458610--2458614.5 and 2459346--2459348: these data present artificial trends due to the momentum dumps of the telescope. We also visually identified and excluded a short bump in the \ac{LC} between BJD 2459337.6 and 2459337.9, most likely an instrumental artifact or some short term photometric variability feature.
The final \acp{LC} cover 7 and 8 transits for sector 11 and 38 respectively.



\subsection{\textit{CHEOPS} observations}\label{sec:CHEOPSobs}

\cheops\ \citep{Benz2021} observed the \wcso\ system during six secondary eclipses of the planet \wcsob\ with a cadence of 60~s. The aim is the measurement of the eclipse depth and derivation of the brightness of the planet in the \cheops\ passband (3500--11000~\AA). Each visit is $\sim$12~hr long, scheduled in order to bracket the eclipse and equally long pre- and post-eclipse photometry. The logbook of the observations, which are part of the \cheops\ Guaranteed Time Observation (GTO) program, is summarized in Table~\ref{tab:obs}.

\begin{table*}
\caption{Logbook of the \cheops\ observations of \wcso. The filekey is the unique identifier associated with each dataset processed by the \cheops\ DRP.}             
\label{tab:obs}      
\centering          
\begin{tabular}{c c c c c c}     
\hline\hline       
Filekey & Start time ]UT] & Visit duration [hr] & Exposure time [s] & N. frames & Efficiency [\%]\\
\hline                    
PR100016\_TG014201\_V0200  &  2021-04-05 12:49:30  &   11.54  &  60.0  &  427  &  61.7 \\
PR100016\_TG014202\_V0200  &  2021-04-15 13:54:09  &   12.14  &  60.0  &  473  &  64.9 \\
PR100016\_TG014203\_V0200  &  2021-05-02 06:56:09  &   11.54  &  60.0  &  465  &  67.2 \\
PR100016\_TG014204\_V0200  &  2021-05-22 08:52:09  &   11.54  &  60.0  &  431  &  62.3 \\
PR100016\_TG014205\_V0200  &  2021-05-28 23:49:09  &   11.44  &  60.0  &  415  &  60.5 \\
PR100016\_TG014206\_V0200  &  2022-05-21 22:33:49  &   11.54  &  60.0  &  445  &  64.3 \\

\hline                  
\end{tabular}
\end{table*}

The data were reduced using version 13 of the \cheops\ \ac{DRP} \citep{Hoyer2020}. This pipeline performs the standard calibration steps (bias, gain, non-linearity, dark current and flat fielding) and corrects for environmental effects (cosmic rays, smearing trails from nearby stars, and background) before the photometric extraction. 

As for the case of \tess\ \acp{LC}, the aperture photometry is significantly contaminated by ASASSN-V J150908.07-424253. To decontaminate the \ac{LC} of \wcso\ we performed the photometric extraction using a modified version of the package PIPE\footnote{\url{https://github.com/alphapsa/PIPE}} \citep{Brandeker2022,Morris2021,Szabo2021}, upgraded in order to compute the simultaneous \ac{PSF} photometry of the target and the background contaminant.

The extracted \acp{LC} present gaps due to Earth occultations which cover $\sim$40\% of the visits. The exposures close to the gaps are characterized by high value of the background flux, due to stray-light from Earth. The corresponding flux measurements are thus affected by a larger photometric scatter. To avoid these low quality data, we applied a 5$\sigma$ clipping to the background measurements: this selection criterion removes less than 20\% of the data with the highest background counts. Finally, for a better outlier rejection, we smoothed the data with a Savitzky-Golay filter, computed the residuals with respect to the filtered \ac{LC} and 5$\sigma$-clipped the outliers. This last rejection criterion excludes a handful of data points in each \ac{LC}.

Finally, we normalize the unsmoothed \acp{LC} by the median value of the photometry. These normalized \acp{LC} are publicly available at CDS {\bf PUT LINK}. 

\section{Stellar radius, mass and age}\label{sec:spectroscopy}

We use a \ac{IRFM} in a \ac{MCMC} approach to  determine the stellar radius of \wcso\ \citep{Blackwell1977,Schanche2020}. We downloaded the broadband fluxes and uncertainties from the most recent data releases for the following bandpasses: {\it Gaia} G, G$_{\rm BP}$, and G$_{\rm RP}$, 2MASS J, H, and K, and {\it WISE}\ W1 and W2 \citep{Skrutskie2006,Wright2010,GaiaCollaboration2021}. Then we matched the observed photometry with synthetic photometry computed in the same bandpasses by using the theoretical stellar \acp{SED} corresponding to stellar atmospheric parameters (Table~\ref{tab:parameters}). The fit is performed in a Bayesian framework and, to account for uncertainties in stellar atmospheric modeling, we averaged the \textsc{atlas} \citep{Kurucz1993,Castelli2003} and \textsc{phoenix} \citep{Allard2014} catalogs to produce weighted averaged posterior distributions. This process yields a $R_{\star}=1.722\pm0.020\, R_{\odot}$.

Assuming the $T_{\mathrm{eff}}$, [Fe/H], and $R_{\star}$ listed in Table~\ref{tab:parameters} as input parameters, we also computed the stellar mass $M_{\star}$ and age $t_{\star}$ by using two different sets of stellar evolutionary models. In detail, we employed the isochrone placement algorithm \citep{bonfanti15,bonfanti16} and its capability of interpolating within pre-computed grids of PARSEC\footnote{\textit{PA}dova and T\textit{R}ieste \textit{S}tellar \textit{E}volutionary \textit{C}ode: \url{http://stev.oapd.inaf.it/cgi-bin/cmd}} v1.2S \citep{marigo17} for retrieving a first pair of mass and age estimates.
A second pair of mass and age values, instead, was computed by CLES \citep[Code Liègeois d'Évolution Stellaire;][]{scuflaire08}, which generates the best-fit evolutionary track of the star by entering the input parameters into the Levenberg-Marquadt minimisation scheme as described in \citet{salmon21}.
After carefully checking the mutual consistency of the two respective pairs of estimates through the $\chi^2$-based criterion broadly discussed in \citet{bonfanti21}, we finally merged the outcome distributions and we obtained $M_{\star}=2.169_{-0.089}^{+0.083}\,M_{\odot}$ and $t_{\star}=50_{-50}^{+60}$ Myr.

\section{Light curve analysis}\label{sec:modeling}

\subsection{\tess\ photometry}\label{sec:tessPer}

\begin{figure*}
    \centering
    \includegraphics[width=0.62\linewidth]{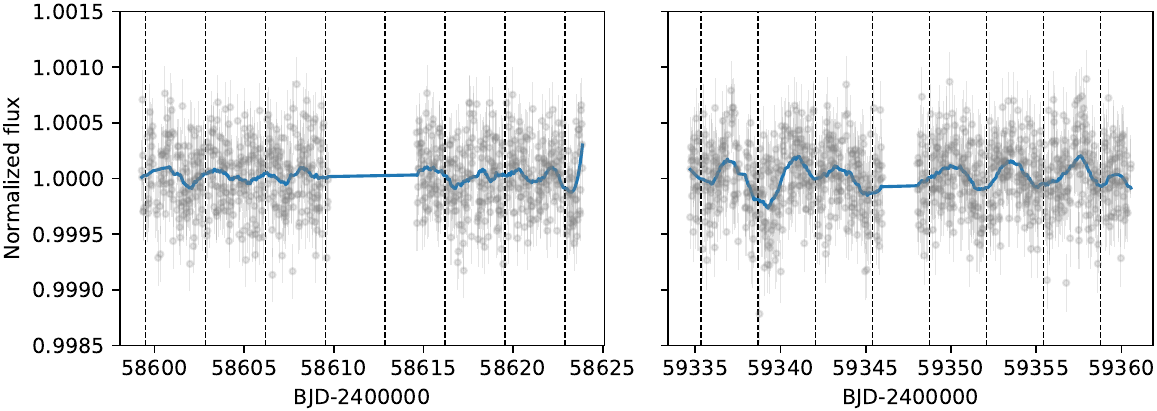}
    \includegraphics[width=0.37\linewidth]{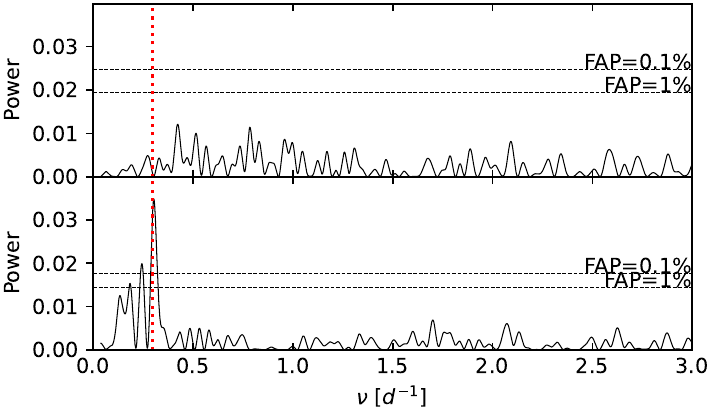}
    \caption{Out-of-transit and out-of-eclipse 30 min cadence \tess\ photometry of \wcso\ during sector 11 (left panel) and sector 38 (central panel). In each panel, the dashed vertical lines mark the planetary transit, while the solid blue line is a smoothing of the data points to emphasize the correlated noise. The right panel shows the GLS periodogram of sector 11 and 38 (top and bottom box respectively). In each box, we report the bootstrap-computed 0.1\% and 1\% \ac{FAP} levels (horizontal dashes) and the planetary orbital period (vertical red dotted line).}\label{fig:periodograms}
\end{figure*}

To compare in a homogeneous way the \acp{LC} of sectors 11 and 38, we rebinned the photometry of sector 38 to 30 min. The standard deviation of the \tess\ \acp{LC}, after transits and secondary eclipses are clipped, is of $\sim$300 ppm for both sectors, and in both cases the photometric uncertainty can account for only $\sim$70\% of the variance. This indicates that there is some noise in the \acp{LC} due to astrophysical signals and/or instrumental leftovers. To investigate if the unexplained variance is related to periodic signals, for both sectors we compute the \ac{GLS} periodogram \citep{Zechmeister2009} of the out-of-transit and out-of-eclipse photometry (Fig.~\ref{fig:periodograms}). For sector 11 we did not find any significant periodic signal, while for sector 38 there is a strong peak at frequency $\rm \nu=0.304\pm0.003~d^{-1}$ with \ac{FAP} lower than 0.1\%. The amplitude of the corresponding sinusoidal signal is $90\pm10$~ppm. 

The periodicity detected in sector 38 is consistent within uncertainties with the planetary orbital period. Nonetheless, we exclude that this signal is of planetary origin because it remained undetected in sector 11 and is not coherent from orbit to orbit in sector 38. This signal might be due to stellar rotation, and is consistent with the typical amplitude reported by \citet{Balona2011} for A-type stars. Its periodicity would correspond to the \vsini\ in Table~\ref{tab:parameters} if the inclination of the stellar rotation axis is $\sim15^{\circ}$. This speculation supports the hypothesis of \citet{Rodriguez2020} that the star is seen nearly pole-on. A significant misalignment between the stellar rotation axis and the planetary orbit axis is not surprising, both because of its young age (no time for realignment to occur), and the stellar temperature. As a matter of fact, hot stars (\teff>6200 K) are observed to often have large misalignments, which is thought to be due to their lack of a convective zone \citep{Winn2010}, needed to tidally align the planetary orbit with the stellar spin axis. Furthermore, \citet{Albrecht2021} found that misaligned orbits are most often polar, or close to polar, as seems to be the case of \wcsob. We expand the discussion on the photometric variability and the orientation of the stellar rotation axis in Appendix~\ref{sec:transitsHighCadence}.


The fact that the period of the correlated noise is similar to the orbital period of \wcsob\ makes it difficult to extract the complete planetary \ac{PC} signal. We thus first tried a simpler and more robust approach to analyze the planetary transits and eclipses (Sect.~\ref{sec:tessTR}), then we attempted a more complex analysis framework aimed at retrieving the full \ac{PC} of \wcsob\ (Sect.~\ref{sec:tessPC}).

\subsubsection{Fit of transits and eclipses}\label{sec:tessTR}

We computed the ephemeris of the planet by trimming segments of the \acp{LC} centered on the transit events (7 transits in sector 11 and 8 transits in sector 8) and as wide as 3 times the transit duration. To further constrain the ephemeris of \wcsob, we included in our analysis the WASP-South photometry (2006 May -- 2014 Aug) and EulerCAM I-band photometry (2018 Mar 26) presented in \citet{Hellier2019}. Also for the case of the WASP-South photometry, we trimmed the \ac{LC} around the transits and kept the 24 intervals containing more than 20 data points.

We fit the data using the same Bayesian approach described in \citet{Scandariato2022}. In summary, it consists in the fit  of a model - in a likelihood-maximization framework - which includes the transits, a linear term for each transit to detrend against stellar/instrumental systematics and a jitter term to fit the white noise not included in the photometric uncertainties. The transit profile is formalized using the quadratic limb darkening (LD) law indicated by \citet{Mandel2002} with the reparametrization of the LD coefficients suggested by \citet{Kipping2013}. For \tess\ sector 11 the model is rebinned to the same 30 min cadence of the data. The likelihood maximization is performed with \ac{MCMC} using the python \texttt{emcee} package version 3.1.3 \citep{Foreman2013}, using a number of samplings long enough to ensure convergence. We used flat priors for all the fitting parameters but the stellar density, for which we used the Gaussian prior $N(0.43,0.02)$ given by the stellar mass and radius in Table~\ref{tab:parameters}. We also used the Gaussian priors for the \ac{LD} coefficients given by the \texttt{LDTk} package \footnote{\url{https://github.com/hpparvi/ldtk}}. For simplicity, we use the same \ac{LD} coefficients for the three datasets. This is motivated by the fact that the WASP-South photometry is not accurate enough to constrain the \ac{LD} profile, and that the \tess\ passband is basically centered on the standard I band, thus the same \ac{LD} profile is expected for the \tess\ and EulerCAM \acp{LC}. Since the model fitting is computationally demanding, we ran the code in the HOTCAT computing infrastructure \citep{Bertocco2020,Taffoni2020}.

\begin{figure*}
    \centering
    \includegraphics[width=.48\linewidth]{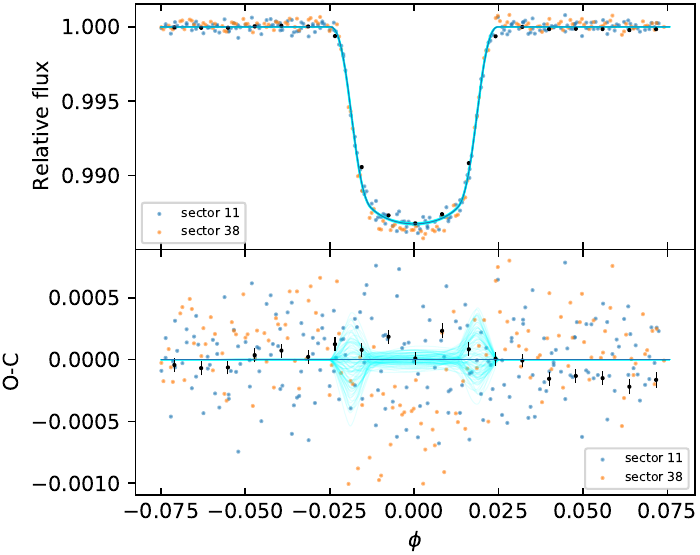}
    \includegraphics[width=.48\linewidth]{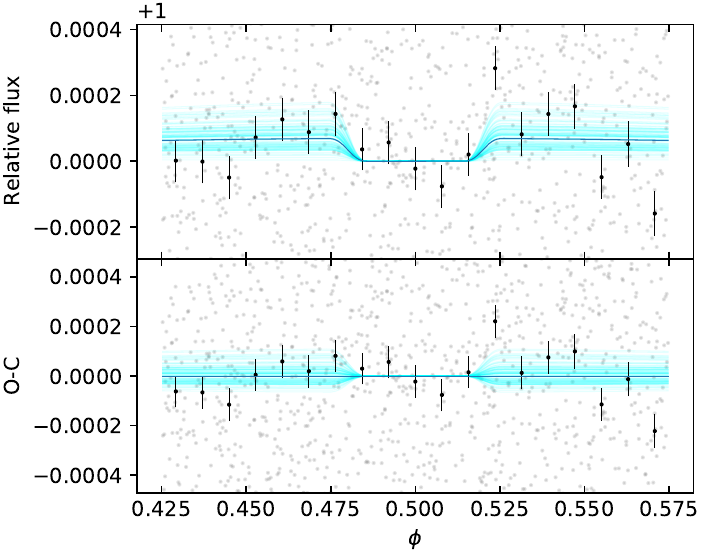}
    \caption{Best fit of the transits and eclipses observed by \tess. \textit{Left - }Detrended and phase-folded planetary transits (top panel). The photometry of sector 11 and 38 are shown with different colors to emphasize the difference in transit depth (the photometry of sectors 11 and 38 is systematically offset respectively upwards and downwards with respect to the bestfit transit profile). The solid blue line is the best fit model, while the cyan lines represent 100 models corresponding to random samples of the \ac{MCMC} fit. For the sake of comparison, the photometry of sector 38 and the theoretical models are rebinned to the same 30 min cadence of sector 11. The black dots represent the rebinned photometry. The corresponding O-C diagram is shown in the bottom panel. \textit{Right - }Same as in the left panel, but centered on the eclipse. For clarity, we do not mark the two sectors with different colors.} \label{fig:phaseFoldedTess}
\end{figure*}

The result of the model fitting is listed in Table~\ref{tab:ephemeris}. The orbital solution we derived is consistent with previous studies \citep{Hellier2019,Rodriguez2020} within 5$\sigma$. The best fit model, corresponding to the \ac{MAP} parameters, is over-plotted to the phase-folded data in Fig~\ref{fig:phaseFoldedTess}, where we rebinned the model and the \ac{LC} of sector 38 to the same 30 min cadence of sector 11 (we do not show the WASP-South and EulerCAM photometry not to clutter the plot). In Fig.~\ref{fig:tessTRcorner} we show the corner plot of the system's parameters from the fit of the transit \acp{LC}.

\begin{table*}
\begin{center}
\caption{Model parameters for the fit of the \tess\, WASP-South and EulerCAM data.}\label{tab:ephemeris}
\begin{tabular}{lllllll}
\hline\hline
Jump parameters & Symbol & Units & \ac{MAP} & C.I.\tablefootmark{a} & Prior\\
\hline
Time of transit & $T_0$ & BJD$_{\rm TDB}$-2400000 & 56612.6581 & 56612.6581(3) & $U$(56612.6, 56612.7)\\
Orbital frequency & $\nu_{\rm orb}$ & days$^{-1}$ & 0.29896856 & 0.29896855(3) & $U$(0.2989,0.2990) \\
Stellar density & $\rho_\star$ & $\rho_\sun$ & 0.45 & 0.44(1) & $N$(0.43,0.02)\\
Radii ratio\tablefootmark{b} & $R_p/R_\star$ & --- & 0.1124 & 0.1125(2) & $U$(0.05,0.12) \\
Radii ratio\tablefootmark{c} & $R_p/R_\star$ & --- & 0.1109 & 0.1108(4) & $U$(0.05,0.12) \\
Radii ratio\tablefootmark{d} & $R_p/R_\star$ & --- & 0.1141 & 0.1141(4) & $U$(0.05,0.12) \\
Impact parameter & b & --- & 0.52 & 0.51(1) & $U$(0.01,0.9) \\
First LD coef.\tablefootmark{b} & q$_{\rm 1}$ & --- & 0.147 & 0.147(8) & $N$(0.133,0.014) \\
Second LD coef.\tablefootmark{b} & q$_{\rm 2}$ & --- & 0.344 & 0.34(2) & $N$(0.333,0.023) \\
Secondary eclipse depth & $\delta_{\rm ecl}$ & ppm & 70 & 70(40) & $U$(0,400)\\
\hline
\hline
Derived parameters & Symbol & Units & \ac{MAP} & C.I. & \\
\hline
Planetary radius\tablefootmark{b} & R$\rm _p$ & R$\rm _J$ & 1.88 & 1.88(2) & including the stellar radius uncertainty\\
Planetary radius\tablefootmark{c} & R$\rm _p$ & R$\rm _J$ & 1.86 & 1.85(2) & including the stellar radius uncertainty\\
Planetary radius\tablefootmark{d} & R$\rm _p$ & R$\rm _J$ & 1.91 & 1.91(2) & including the stellar radius uncertainty\\
Orbital period & P$_{\rm orb}$ & day & 3.3448332 & 3.3448333(4) & \\
Transit duration & T$_{\rm 14}$ & hr & 3.488 & 3.488(8) & \\
Scaled semi-major axis & $a/R_\star$ & --- & 7.20 & 7.19(6) & \\
Orbital inclination & $i$ & degrees & 85.8 & 85.8(1) & \\
\hline
\end{tabular}
\tablefoot{
        \tablefoottext{a}{Uncertainties expressed in parentheses refer to the last digit(s).}
        \tablefoottext{b}{Fitting WASP-South, EulerCAM and \tess\ data altogether.}
        \tablefoottext{c}{Fitting \tess\ sector 11 only.}
        \tablefoottext{d}{Fitting \tess\ sector 38 only.}
}
\end{center}
\end{table*}

Since the \tess\ \acp{LC} of sector 11 and 38 show different levels of variability, we investigated any seasonal dependence of the apparent planet-to-star radius ratio. We used the same Bayesian framework as above, where we fix the ephemeris of \wcso. The planet-to-star radius ratio we derived is 0.1108$\pm$0.0004 for sector 11 and 0.1141$\pm$0.0004 for sector 38. The difference is thus 0.0033$\pm$0.0006, that is we found different transit depths with a 5.5$\sigma$ significance. In particular, we remark that the planet looks larger in sector 38, where the rotation signal is stronger. We thus speculate that \wcso\ was in a low variability state during sector 11, while one year later ($\sim$80 stellar rotations) during sector 38 the stellar surface hosted dark spots in co-rotation with the star. This hypothesis is consistent with \citet{Hummerich2018}, according to which magnetic chemically peculiar stars may show complex photometric patterns due to surface inhomogeneities.

We used a similar framework to extract the secondary eclipse signal of \wcso: we simultaneously fit segments of the \tess\ \acp{LC} centered on the planetary eclipses keeping fixed the ephemeris of \wcso\ to the values listed in Table~\ref{tab:ephemeris}. We did not attempt the disjoint analysis of the two sectors as the expected eclipse depth is of the order of $\sim$100~ppm (see below), and the photometric precision of the \acp{LC} is not good enough to appreciate differences in the eclipse depth with enough statistical evidence. The only free planetary parameter of the model is thus the eclipse depth, which turned out to be $\delta_{\rm ecl}=70\pm40$~ppm (MAP: 70 ppm). The detrended phase-folded eclipses are shown in Fig~\ref{fig:phaseFoldedTess} together with the \ac{MAP} model.

\subsubsection{Fit of the phase curve}\label{sec:tessPC}

As a more advanced data analysis, we jointly fit the two \tess\ sectors to extract the full planetary \ac{PC} (for a homogeneous analysis, we rebinned sector 38 to the same 30 min cadence of sector 11). The fitting model now includes the transits, secondary eclipses, the planetary \ac{PC}, a \ac{GP} to fit the correlated noise in the data and, for each \tess\ sector, a long-term linear trend and a jitter term. Given the system parameters listed in Table~\ref{tab:parameters}, the expected amplitude of ellipsoidal variations \citep{Morris1985} and Doppler boosting \citep{Barclay2012} in the planetary \ac{PC} are 2.5~ppm and 1~ppm respectively, out of reach for the \tess\ photometry. Hence, for the sake of simplicity, we do not include them in the model fitting.

We jointly fit the two \tess\ sectors, starting with the simplest model where the out of transit \ac{PC} is flat (that is, we assumed that the planetary \ac{PC} is not detectable) and the putative stellar rotation is modeled as \ac{GP} with the SHO kernel\footnote{Provided by the \texttt{celerite2} python package version 0.2.1 \citep{Foreman2017,Foreman2018}}, indicated for quasi-periodic signals. 
Checking the residuals of the fit we notice that some aperiodic correlated noise is left. We thus increased the complexity of the \ac{GP} model by adding a Mat\'ern3/2 kernel to capture the remaining correlated noise adding only two free parameters to the model. 
This composite GP model turned out to ensure better convergence to the fit of the model.

Finally, we also included the planetary \ac{PC}. Assuming that the planet is basically composed of two homogeneous dayside and nightside, the \ac{PC} is in principle the combination of three components: the \ac{PC} due to reflected light (with amplitude $A_{\rm refl}$), the \ac{PC} of the planetary night side (with amplitude $A_{\rm n}$) and the \ac{PC} of the planetary day side, whose amplitude $A_{\rm d}$ is parameterized such to be larger than $A_{\rm n}$ by an increment $\delta_{\rm d}$ (this parametrization avoids the nonphysical case of a planetary night side brighter than the dayside). The three terms are modeled assuming Lambert's cosine law.

For the reflected \ac{PC}, the amplitude can be expressed as:
\begin{equation}
A_{\rm refl}=A_{\rm g}\left(\frac{R_{\rm p}}{a}\right)^2,\label{eq:albedo}
\end{equation}
where \ageo\ is the planetary geometric albedo \citep{Seager2010}. Assuming a Lambertian reflective planetary surface, the geometric albedo is fixed by the Bond albedo $A_{\rm B}$ by \ageo=$\frac{2}{3}A_{\rm B}$. In the optimistic case of a perfectly reflecting body ($A_{\rm B}$=1), the maximum amplitude for the reflected \ac{PC} is thus $A_{\rm refl}=167\pm5$~ppm, where we have used the system parameters in Table~\ref{tab:ephemeris}.

With a slight re-adaptation of the formalism in \citet{Seager2010}, the amplitude thermal component $A_{\rm therm}$ can be expressed as:
\begin{equation}
A_{\rm therm}=\left(\frac{R_{\rm p}}{R_\star}\right)^2\frac{\int\eta(\lambda) B(\lambda,T)d\lambda}{\int\eta(\lambda) I_\star(\lambda,T_{\rm eff})d\lambda},\label{eq:thermal}
\end{equation}
where $\eta(\lambda)$ is the optical throughput of the telescope and $I_\star(\lambda,T_{\rm eff})$ is the expected stellar intensity computed with the NextGen model \citep{Hauschildt1999} corresponding to the effective temperature $T_{\rm eff}$. For the planetary thermal emission we lack the infrared information which can constrain the emission spectrum. We thus assume a black body spectrum $B(\lambda,T)$ at temperature $T$.

The amplitude of the thermal emission from the dayside and the nightside depends on the respective temperatures, which we estimated following \citet{Cowan2011}. The expected temperature of the substellar point of \wcsob\ is:
\begin{equation}
    T_{\rm 0}=T_{\rm eff}\sqrt{\frac{R_\star}{a}}=3530\pm60~{\rm K},
\end{equation}
where we used the stellar effective temperature in Table~\ref{tab:parameters} and the scaled semi-major axis in Table~\ref{tab:ephemeris}.

The nightside temperature $T_{\rm n}$ and the dayside temperature $T_{\rm d}$ depend on the Bond albedo $A_B$ and heat recirculation coefficient $\epsilon$:
\begin{eqnarray}
    T_{\rm n}&=T_{\rm 0}\left(1-A_{\rm B}\right)^{1/4}\left(\frac{\epsilon}{4}\right)^{1/4}\label{eq:tn}\\
    T_{\rm d}&=T_{\rm 0}\left(1-A_{\rm B}\right)^{1/4}\left(\frac{2}{3}-\frac{5}{12}\epsilon\right)^{1/4}\label{eq:td}
\end{eqnarray}

The highest dayside temperature, corresponding to $A_{\rm B}$=0 and no heat recirculation ($\epsilon=0$), is:
\begin{equation}
    T_{\rm d}=T_{\rm 0}\left(\frac{2}{3}\right)^{1/4}=3190\pm60~{\rm K},
\end{equation}
while the maximum nightside temperature, obtained in the case of full energy recirculation ($\epsilon=1$), is:
\begin{equation}
    T_{\rm n}=T_{\rm 0}\left(\frac{1}{4}\right)^{1/4}=2500\pm40~{\rm K}.
\end{equation}
These maximum temperatures lead to the upper limit on the dayside ($A_{\rm d}$) and nightside ($A_{\rm n}$) thermal emission amplitude of:
\begin{equation}
    A_{\rm d}=225\pm27~{\rm ppm}; A_{\rm n}=48\pm7~{\rm ppm}.
\end{equation}

\begin{figure}
    \centering
    \includegraphics[width=\linewidth]{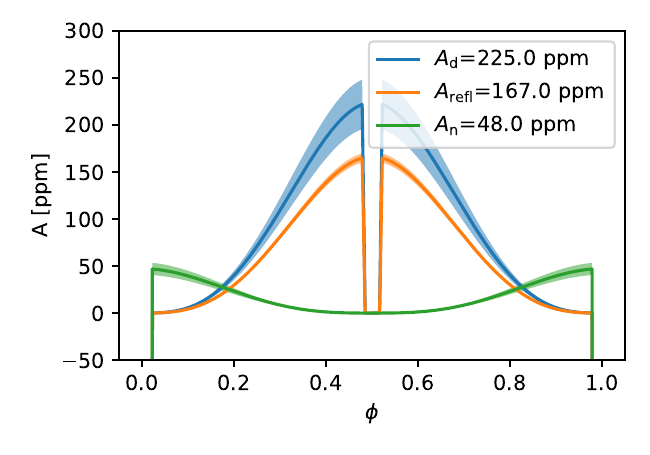}
    \caption{Upper limits on the reflected \ac{PC}, dayside thermal \ac{PC} and nightside thermal \ac{PC} for \wcsob.}\label{fig:phaseCurves}
\end{figure}

In Fig.~\ref{fig:phaseCurves} we plot the upper limits for the three \acp{PC} discussed so far. The figure shows that the three components may reach amplitudes of comparable orders of magnitude, meaning that none of them can in principle be neglected in the extraction of the planetary PC. We note that the two components belonging to the dayside (reflection and thermal emission) have the same shape, thus it is not possible to disentangle them. To simplify the model and avoid the degeneracy between dayside reflection and emission, we thus artificially set $A_{\rm refl}=0$ and let $A_{\rm d}$ absorb the whole signal belonging to the planetary dayside.

The Akaike Information Criterion \citep[AIC,][]{Burnham2002} favors the model that includes the planetary \acp{PC}. The simpler model with flat out-of-transit \ac{PC} has a relative likelihood of 55\% and cannot be rejected. The posterior distributions of the common parameters (system ephemeris and planet-to-star radii ratio) between the two models do not differ significantly.

For both models, the SHO \ac{GP} has a frequency $\nu\sim$0.30 d$^{-1}$, an amplitude $\sim$70~ppm and a timescale $\sim$11~d. Frequency $\nu$ and the amplitude of the quasi-periodic \ac{GP} are consistent with the frequency and amplitude of the periodic signal found in the periodogram (Sect.~\ref{sec:tessPer}). While the correspondence between the orbital period of \wcsob\ and the periodicity of the \ac{GP} suggests that the red noise in the data is of planetary origin, we do not have other strong evidence that support this hypothesis. On the contrary, the planetary origin seems odd for the reasons discussed in Sect.~\ref{sec:tessPer}. The aperiodic red noise fitted as a Mat\'ern3/2 \ac{GP} has an amplitude $\sim$150~ppm and a timescale of $\sim$50~min. Despite the aperiodic red noise evolves on timescales shorter than the transit duration ($T_{\rm 14}$=3.53~hr), the retrieved posterior distributions of the orbital parameters are similar to the ones obtained in Sect.~\ref{sec:tessTR}.

Regarding the retrieval of the planetary \acp{PC}, we derived an upper limit of $A_{\rm n}<$174~ppm at the 99.9\% confidence level, while for the dayside we derived a 1$\sigma$ confidence band of 110$\pm$40~ppm which includes both the reflection and thermal emission components. We remark that the $A_{\rm d}$ parameter corresponds in our model to the eclipse depth $\delta_{\rm ecl}$ and is consistent within uncertainties with the estimate reported in Table~\ref{tab:ephemeris}.

The fit of the full \tess\ \ac{LC} thus confirms the results previously obtained in Sect.~\ref{sec:tessTR}. Given the similarities between the results obtained with the two approaches, we preferred the former one as it is less prone to interference between the data detrending and the extraction of the orbital parameters and the eclipse depth.

\subsection{\cheops\ photometry}\label{sec:CHEOPSfit}

To analyze the \cheops\ \acp{LC} we use the same approach as for the fit of the eclipses observed by \tess\ (Sect.~\ref{sec:tessTR}), with an additional module in the fitting model that takes into account the systematics in the \cheops\ data. As a matter of fact, \cheops\ photometry is affected by variable contamination from background stars in the field of view \citep[e.g. ][]{Lendl2020,Deline2022,Hooton2021,Wilson2022,Scandariato2022}. This variability is due to the interplay of the asymmetric \ac{PSF} with the rotation of the field of view \citep{Benz2021}. PIPE by design uses nominal magnitudes and coordinates of the stars in the field to fit the stellar \acp{PSF}, but residual correlated noise is present in the \acp{LC} due to inaccuracies in the assumptions.
This signal is phased with the roll angle of the \cheops\ satellite and, in the case of \wcso, is clearly visible in the \ac{GLS} periodogram of the data together with its harmonics. To remove this signal, we included in our algorithm a module which fits independently for each visit the harmonic expansions of the telecope's orbital period and its harmonics.
A posteriori, we found that the roll angle phased modulation is adequately suppressed (that is, there is no peak in the periodograms at the frequencies in the harmonic series) if we include in the model the fundamental harmonic and its first two harmonics.

We also noticed that the photometry is significantly correlated with the coordinates of the centroid of the \ac{PSF} on the detector. To decorrelate against this instrumental jitter we thus included in the model a bi-linear function of the centroid coordinates.

Finally, to take into account a white noise not included in the formal photometric uncertainties, we add to our model a diagonal \ac{GP} kernel of the form:
\begin{equation}
k(t_i,t_j)=j_v^2\delta_{i,j},\label{eq:kernelCHEOPS}
\end{equation}
with an independent jitter term $j_v$ for each CHEOPS visit.

\begin{figure}
    \centering
    \includegraphics[width=\linewidth]{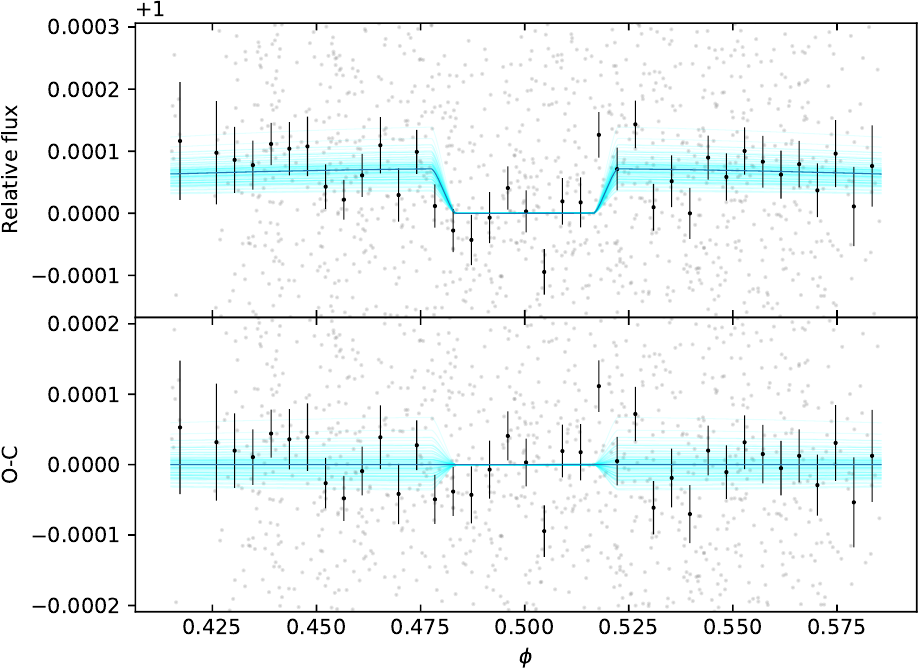}
    \caption{Same as the right panel in Fig.~\ref{fig:phaseFoldedTess} for the eclipses of \wcso\ observed by \cheops.} \label{fig:phaseFoldedCheops}
\end{figure}

As for the fit of the \tess\ data (Sect.~\ref{sec:tessTR}), we search the bestfit parameters through likelihood-maximization in a \ac{MCMC} framework, and we find $\rm \delta_{ecl}=70\pm20$~ppm. The ephemeris of the planet were locked to the \ac{MAP} values listed in Table~\ref{tab:ephemeris}, which guarantee an uncertainty on the eclipse time less than 23 s for each CHEOPS visit. The phase-folded data detrended against stellar and instrumental correlated noise are shown in Fig.~\ref{fig:phaseFoldedCheops}.

\section{Discussion}\label{sec:discussion}

In the previous sections we have analyzed the \tess\ and \cheops\ \acp{PC} of \wcsob\ in order to extract its planetary eclipse depth $\delta_{\rm ecl}$, obtaining respectively $70\pm40$~ppm and $70\pm20$~ppm. These two measurements take into account both the reflection from the planetary dayside and its thermal emission. With the assumptions discussed below, the combined information on the eclipse depth from two different instruments allows to disentangle these two contributions. For a given telescope with passband $\eta(\lambda)$ we can combine Eq.~\ref{eq:albedo} and Eq.~\ref{eq:thermal} obtaining:
\begin{equation}
    \delta_{\rm ecl}=A_{\rm g}\left(\frac{R_{\rm p}}{a}\right)^2+\left(\frac{R_{\rm p}}{R_\star}\right)^2\frac{\int\eta(\lambda) B(\lambda,T_{\rm d})d\lambda}{\int\eta(\lambda) I_\star(\lambda,T_{\rm eff})d\lambda},\label{eq:decl}
\end{equation}
where $T_{\rm d}$ is the planetary dayside temperature.

Solving for \ageo, and indicating with the subscripts $C$ (for \cheops) and $T$ (for \tess) the wavelength-dependent quantities, we obtain:
\begin{equation}
\begin{cases}
    A_{\rm g}^C=\left(\frac{a}{R_\star}\right)^2\left[\frac{\delta_{\rm ecl}^C}{\left(\frac{R_{\rm p}}{R_\star}\right)^2}-\frac{\int\eta^C(\lambda) B(\lambda,T_{\rm d})d\lambda}{\int\eta^C(\lambda) I_\star(\lambda,T_{\rm eff})d\lambda}\right],\\
    A_{\rm g}^T=\left(\frac{a}{R_\star}\right)^2\left[\frac{\delta_{\rm ecl}^T}{\left(\frac{R_{\rm p}}{R_\star}\right)^2}-\frac{\int\eta^T(\lambda) B(\lambda,T_{\rm d})d\lambda}{\int\eta^T(\lambda) I_\star(\lambda,T_{\rm eff})d\lambda}\right].\label{eq:ag}   
\end{cases}
\end{equation}

In the most general case, $A_g^C$ and $A_g^T$ differ according to the planetary reflection spectrum. Since we do not have any spectroscopic analysis of the dayside of \wcsob, we define the proportionality coefficient $\alpha=A_g^T/A_g^C$ to account for differences between the two passbands. Solving Eq.~\ref{eq:ag} for $T_{\rm d}$ we thus obtain the implicit function:
\begin{equation}
    \frac{\alpha\delta_{\rm ecl}^C-\delta_{\rm ecl}^T}{\left(\frac{R_{\rm p}}{R_\star}\right)^2}+\frac{\int\eta^T(\lambda) B(\lambda,T_{\rm d})d\lambda}{\int\eta^T(\lambda) I_\star(\lambda)d\lambda}-\alpha \frac{\int\eta^C(\lambda) B(\lambda,T_{\rm d})d\lambda}{\int\eta^C(\lambda) I_\star(\lambda)d\lambda}=0.\label{eq:tdRoot}
\end{equation}

We initially assumed the simplest scenario of a gray albedo spectrum ($\alpha$=1) in the spectral range covered by \cheops\ and \tess\ (the respective bandpasses are shown in Fig.~\ref{fig:stellarSpectrum}). In this scenario, we expect that the contribution of reflection to the eclipse depth is the same in the \cheops\ and \tess\ passbands. We also expect that the thermal emission from a $T\simeq$3000~K black body has a larger contribution in the \tess\ passband, because it favors redder wavelengths compared with \cheops\ (see Fig.~\ref{fig:stellarSpectrum}).

To estimate the dayside temperature $T_{\rm d}$ that best explains our best observations together with its corresponding uncertainty, we randomly extracted 10\,000 steps from the Monte Carlo chains obtained in the fit of the transit and eclipse \acp{LC} (Sect.~\ref{sec:tessTR} and \ref{sec:CHEOPSfit}). We also generated 10\,000 samples of the stellar effective temperature using a normal distribution with mean and standard deviations as reported in Table~\ref{tab:parameters}. For each of the 10\,000 samples we thus numerically solved Eq.~\ref{eq:tdRoot} using the \texttt{fsolve} method of the \texttt{scipy} Python package, thus obtaining 10\,000 samples of $T_{\rm d}$. Then, we plugged  all the samples back into Eq.~\ref{eq:ag} to derive $A_g^C$(=$A_g^T$).

\begin{figure}
    \centering
    \includegraphics[width=\linewidth]{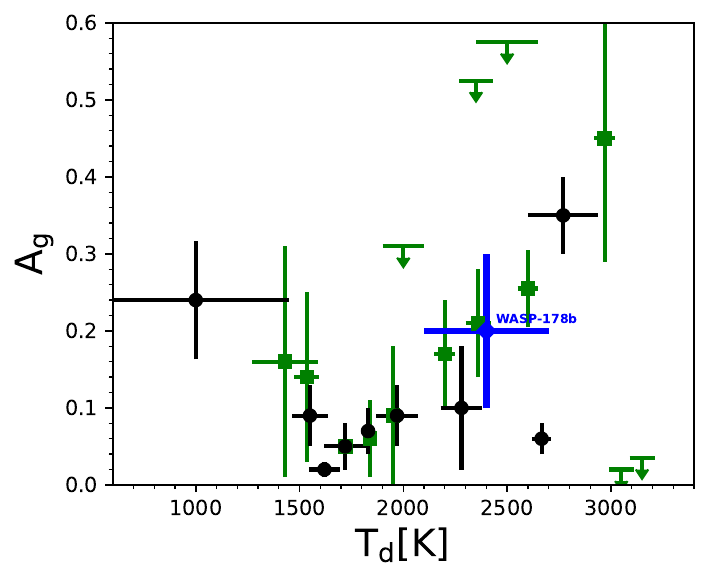}
    \caption{Adaptation of Fig.~10 in \citet{Wong2021} including our analysis of \wcsob\ (in blue). The green squares indicate the systems from the first and second year of the \tess\ primary mission. The black circles indicate the Kepler-/CoRoT-band geometric albedos for the targets that were observed by those missions.}\label{fig:wongPlot}
\end{figure}

\begin{figure}
    \centering
    \includegraphics[width=\linewidth]{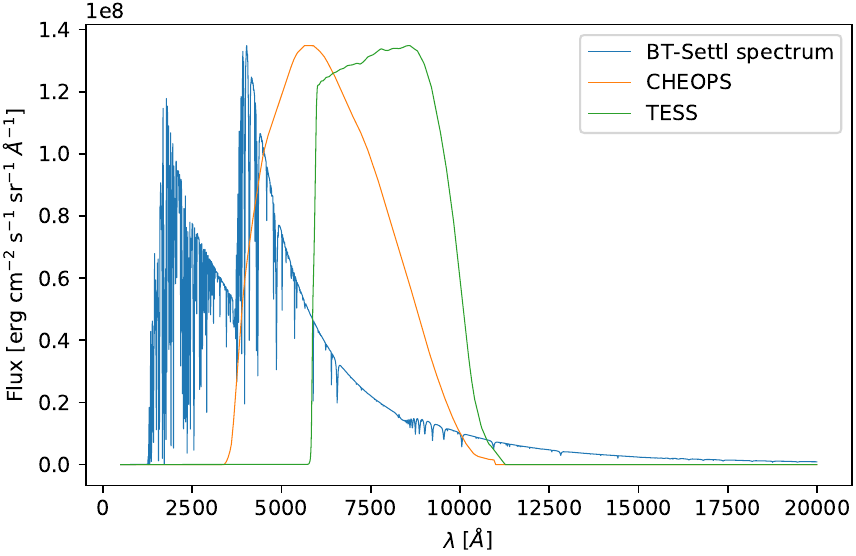}
    \caption{BT-Settl synthetic stellar spectrum for \wcso\ together with the \cheops\ and \tess\ passbands.}\label{fig:stellarSpectrum}
\end{figure}

\begin{figure*}
    \centering
    \includegraphics[width=.49\linewidth]{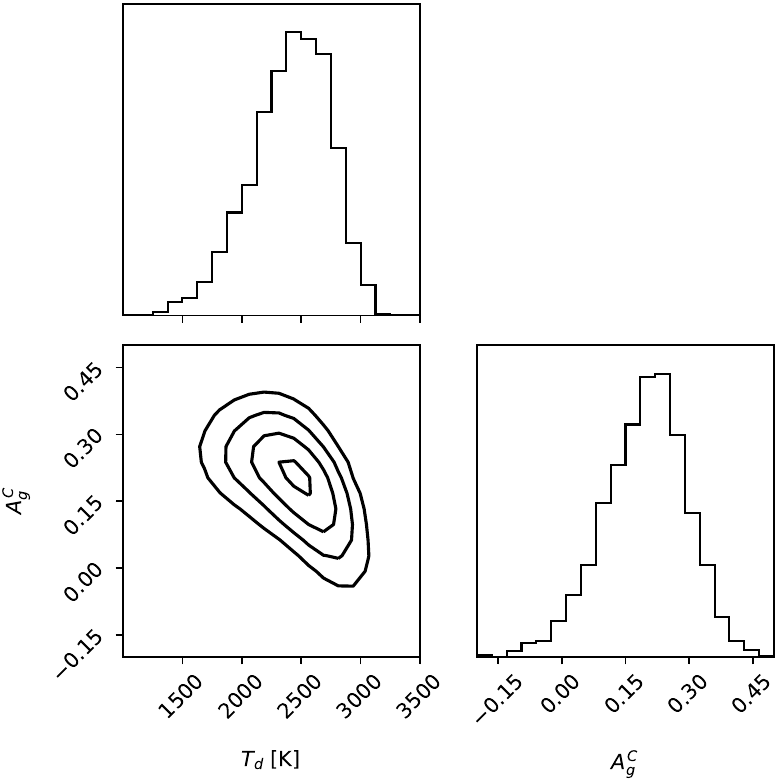}
    \includegraphics[width=.49\linewidth]{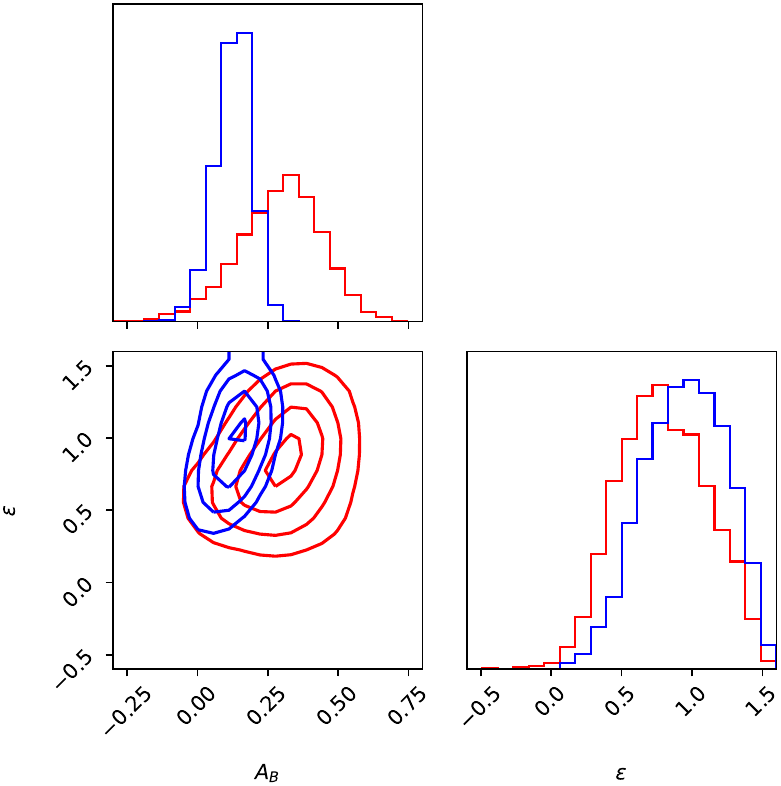}
    \caption{\textit{Left panel - } $A_g^C$--$T_{\rm d}$ density maps in the case of $\alpha=1$. \textit{Right panel - } Albedo--recirculation density maps assuming maximum $A_{\rm B}$ (red contours) and minimum $A_{\rm B}$ (blue contours).}\label{fig:agTd_gray}
\end{figure*}

The root finding algorithm failed for $\sim$50\% of the samples. These failures correspond to the combination of parameters that prevent Eq.~\ref{eq:tdRoot} from having a $T_{\rm d}$ root in the domain of real numbers. This happens in particular for the samples where $\delta_{\rm ecl}^T<\delta_{\rm ecl}^C$ that, as discussed above, are not consistent with the gray albedo hypothesis. We thus reject these samples and show in Fig.~\ref{fig:agTd_gray} the $A_g^C$--$T_{\rm d}$ density map of the remaining ones. Our results indicate $A_g^C=0.2\pm0.1$ and $T_{\rm d}=2400\pm300$~K, {\bf consistent with the temperature-pressure profile derived by \citep{Lothringer2022} at pressure higher that 1$\rm\mu$bar by means of ultraviolet transmission spectroscopy}. Our result also follows the general trend of \ageo\ increasing with $T_{\rm d}$ indicated by \citet{Wong2021} (Fig.~\ref{fig:wongPlot}).

By definition, the geometric albedo \ageo\ refers to the incident light reflected back to the star at a given wavelength (or bandpass). Integrating at all angles, the spherical albedo \asph\ is related to \ageo\ by the \textit{phase integral} $q$: \asph=$q$\ageo\ \citep[see for example][]{Seager2010}. Depending on the scattering law, exoplanetary atmospheres have 1<$q$<1.5 \citep{Pollack1986,Burrows2010}. Unfortunately, for the reasons explained in Sect.~\ref{sec:tessPC}, we could not extract a robust phase curve for \wcsob, hence we could not place any better constraint on $q$. In the following we thus consider the two limiting scenarios $A_S^{\rm min}=A_g$ and $A_S^{\rm max}=1.5A_g$.

The Bond albedo \abond\ is computed as the average of \asph\ weighted over the incident stellar spectrum:
\begin{equation}
    A_{\rm B}=\frac{\int_0^\infty A_{\rm S}(\lambda)I_\star(\lambda)d\lambda}{\int_0^\infty I_\star(\lambda)d\lambda}.\label{eq:abond}
\end{equation}
The conversion into \abond\ thus relies on the measurement of \asph\ across the stellar spectrum. In this study we only covered the optical and near-infrared part of the spectrum, as shown in Fig.~\ref{fig:stellarSpectrum}, and we lack the necessary information in the UV, mid, and far infrared domains. Following the approach of \citet{Schwartz2015}, we explore the scenario of minimum Bond albedo $A_{\rm B}^{\rm min}$ obtained through Eq.~\ref{eq:abond} assuming \asph=$A_{\rm S}^{\rm min}$ in the spectral range covered by \cheops\ and \tess\ and \asph=0 otherwise. The opposite limiting case assumes \asph$(\lambda)=A_{\rm S}^{\rm max}$ at all wavelength, which leads to $A_{\rm B}^{\rm max}=A_{\rm S}^{\rm max}$. To compute the integrals in Eq.~\ref{eq:abond}, we used the synthetic spectrum in the BT-Settl library corresponding to the parameters of \wcso\ \citep{Allard2012}.

Under the hypothesis that $\alpha=1$, we obtained samples of $A_{\rm g}^C=A_{\rm g}^T=$\ageo, which now translate into samples of $A_{\rm B}^{\rm min}$ and $A_{\rm B}^{\rm max}$ with the assumptions discussed above. Moreover, inverting Eq.~\ref{eq:td} yields to:
\begin{equation}
    \epsilon=\frac{1}{5}\left[8-\left(\frac{T_{\rm d}}{T_{\rm eff}}\right)^4\left(\frac{a}{R_\star}\right)^2\frac{12}{1-A_{\rm B}}\right].\label{eq:epsilon}
\end{equation}
Equation \ref{eq:epsilon} indicates that $\epsilon$ is a decreasing function of \abond. Plugging in the samples of $T_{\rm d}$, \teff, $a/R_{\star}$ and alternatively $A_{\rm B}^{\rm min}$ and $A_{\rm B}^{\rm max}$, we obtained the corresponding samples $\epsilon^{\rm max}$ and $\epsilon^{\rm min}$.

The albedo--recirculation density maps of the two scenarios are shown in the right panel of Fig.~\ref{fig:agTd_gray}. Unsurprisingly, we find that the upper limit on \abond\ is larger ($\sim$0.6) in the scenario where all the factors concur to push up the reflectivity of the atmosphere (high $q$ and maximum \asph), and it decreases to $\sim$0.3 in the opposite scenario of minimum reflectivity.

The recirculation coefficient $\epsilon$ tends towards large values, eventually exceeding 1 due to measurement uncertainties, in both scenarios. While $\epsilon>1$ is physically meaningless, still the posterior distribution indicates a high level of atmospheric energy recirculation in both cases of minimum and maximum albedo ($\epsilon^{\rm max}=1.0\pm0.3$ and $\epsilon^{\rm min}=0.8\pm0.3$ respectively). Following different atmospheric models \citep[e.g., ][]{Perez2013,Komacek2016,Schwartz2017,Parmentier2018}, \wcsob\ is in a temperature regime where heat recirculation of zonal winds is suppressed and recirculation efficiency is expected to be low. {\bf Nonetheless, \citet{Zhang2018} collected observational evidence that \acp{HJ} with irradiation temperatures similar to the one of \wcsob\ are characterized by efficient day-to-night recirculation. Using the global circulation models of \citet{Kataria2016}, they ascribed this efficiency to the presence of zonal winds, that also explain the eastward offset of the \ac{PC} of the same planets. While the theoretical predictions on the offset (and the recirculation efficiency correspondingly) likely overestimate the truth \citep[see Fig.~15 in][]{Zhang2018}, still there is an indication that zonal winds might indeed explain the energy transfer from the dayside to the nightside of \wcsob.

} Unfortunately, as discussed in Sect.\ref{sec:tessPC}, it is difficult to disentangle the planetary phase curve and the stellar variability. By consequence, any measurement of the phase curve offset is precluded. Another possibility is that ionized winds flow from the dayside, where temperatures higher that 2500~K leads to complete dissociation of H$_2$, to the nightside, where lower temperatures allow the molecular recombination and the consequent energy release. This scenario is supported by several recent studies \citep{Bell2018,Tan2019,Mansfield2020,Helling2021,Helling2023}. 

\begin{figure*}
    \centering
    \includegraphics[width=.49\linewidth]{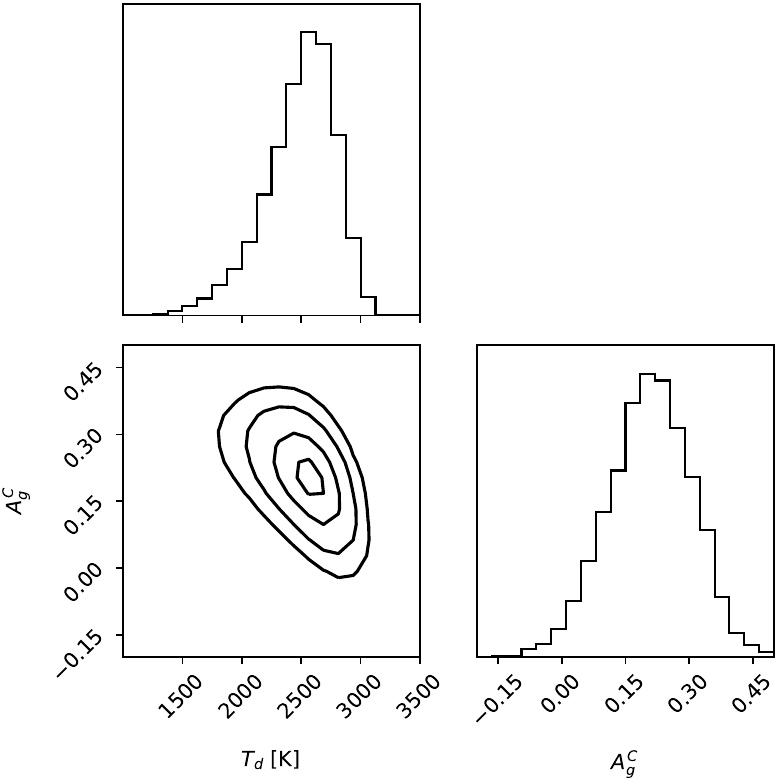}
    \includegraphics[width=.49\linewidth]{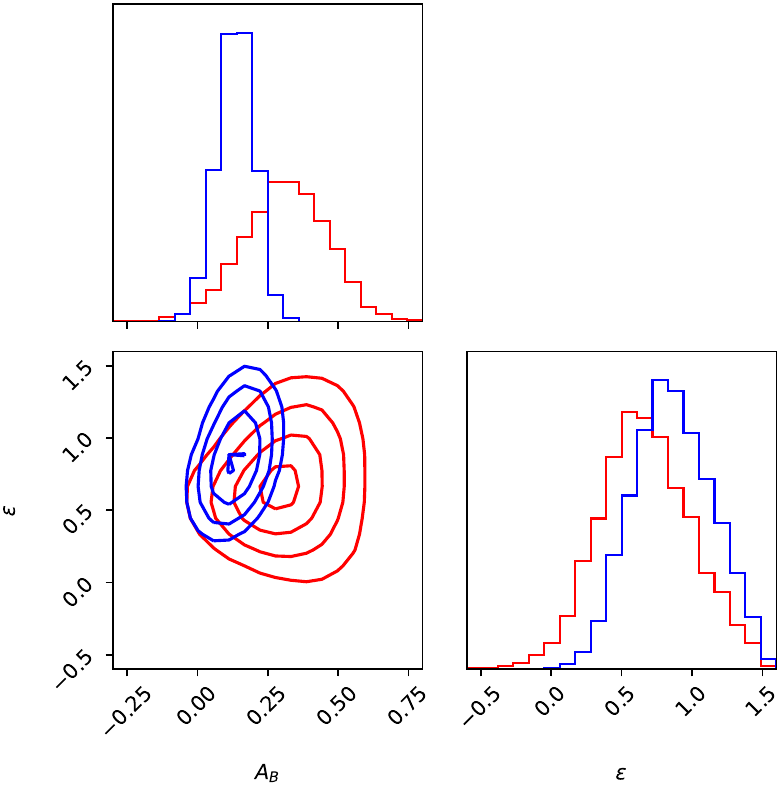}
    \caption{Same as in Fig.~\ref{fig:agTd_gray} in the case $\alpha=0.5$.}\label{fig:agTd_nongray}
\end{figure*}

According to the synthetic models computed by \citet{Sudarsky2000}, the albedo spectrum of highly irradiated giant planets is expected to show molecular absorption bands by H$_2$O at around 1~$\mu m$ and longer wavelength. Since the passband of \tess\ is more sensitive in the near-infrared than \cheops, it is plausible to assume that the geometric albedo in the \tess\ passband is lower than for \cheops. In order to assess how the assumption on $\alpha$ affects our results, we re-run our analysis assuming an extreme $\alpha=0.5$. The results are shown in Fig.~\ref{fig:agTd_nongray}: we find almost the same posterior distribution for dayside temperature, albedo and recirculation, indicating that the most important source of uncertainties are not the assumptions we make but the measurement uncertainties on the eclipse depth extracted from the \cheops\ and \tess\ \acp{LC}.

\section{Summary and conclusion}

In this work we have analyzed the space-borne photometry of the \wcso\ system obtained with \tess\ and \cheops. For both telescopes we have tailored the photometric extraction in order to avoid the strong contamination by a background eclipsing binary. We have found evidence that the stellar host is rotating with a period of $\sim$3~days, consistent with the orbital period of its \ac{HJ}, and that it is seen in a pole-on geometry.

The similarity between the stellar rotation period and the orbital period of the planet, coupled with the quality of the \tess\ data, does not allow a robust analysis of the full \ac{PC} of the planet, nor we could put strong constraints on the planetary nightside emission. Nonetheless, focusing on the transit and eclipse events, we could update the ephemeris of the planet and measure an eclipse depth of 70$\pm$40 ppm and 70$\pm$20~ppm in the \tess\ and \cheops\ passband respectively.

The joint analysis of the eclipse depth measured in the two passbands allowed us to constrain the temperature in the 2250--2750~K range, consistent with the temperature-pressure profile derived by \citep{Lothringer2022} at pressure higher that 1$\rm\mu$bar. Moreover, we could also constrain the optical geometric albedo \ageo<0.4, which fits the general increasing albedo with stellar irradiation indicated by \citet{Wong2021}. 

Finally, we found indication of an efficient atmospheric heat recirculation of \wcsob. If confirmed, this evidence challenges the models that predict a decreasing heat transport by zonal winds as the equilibrium temperature increases \citep[e.g., ][]{Perez2013,Komacek2016,Schwartz2017,Parmentier2018}. Conversely, \wcsob\ is an interesting target to test atmospheric models where heat transport is granted by ionized winds from the dayside to the nightside, where the recombination of H$_2$ takes place \citep[e.g. ][]{Mansfield2020,Helling2021,Helling2023}. Additional observations of \wcsob\ are needed to better constrain its atmospheric recirculation and rank current competing models. The \wcso\ system is going to be observed again by \tess\ in May 2023, but this is not expected to add much to the two sectors analyzed in this work. Conversely, dedicated observations from larger space observatories might provide helpful insights of the physical and chemical equilibrium of the atmosphere of \wcsob.

\begin{acknowledgements}
We thank the referee, N. B. Cowan, for his valuable comments and suggestions. CHEOPS is an ESA mission in partnership with Switzerland with important contributions to the payload and the ground segment from Austria, Belgium, France, Germany, Hungary, Italy, Portugal, Spain, Sweden, and the United Kingdom. The CHEOPS Consortium would like to gratefully acknowledge the support received by all the agencies, offices, universities, and industries involved. Their flexibility and willingness to explore new approaches were essential to the success of this mission. 
IPa, GSc, VSi, LBo, GBr, VNa, GPi, and RRa acknowledge support from CHEOPS ASI-INAF agreement n. 2019-29-HH.0. 
ML acknowledges support of the Swiss National Science Foundation under grant number PCEFP2\_194576. 
This work was also partially supported by a grant from the Simons Foundation (PI Queloz, grant number 327127). 
S.G.S. acknowledge support from FCT through FCT contract nr. CEECIND/00826/2018 and POPH/FSE (EC). 
ABr was supported by the SNSA. 
ACCa and TWi acknowledge support from STFC consolidated grant numbers ST/R000824/1 and ST/V000861/1, and UKSA grant number ST/R003203/1. 
V.V.G. is an F.R.S-FNRS Research Associate. 
YAl acknowledges support from the Swiss National Science Foundation (SNSF) under grant 200020\_192038. 
We acknowledge support from the Spanish Ministry of Science and Innovation and the European Regional Development Fund through grants ESP2016-80435-C2-1-R, ESP2016-80435-C2-2-R, PGC2018-098153-B-C33, PGC2018-098153-B-C31, ESP2017-87676-C5-1-R, MDM-2017-0737 Unidad de Excelencia Maria de Maeztu-Centro de Astrobiologí­a (INTA-CSIC), as well as the support of the Generalitat de Catalunya/CERCA programme. The MOC activities have been supported by the ESA contract No. 4000124370. 
S.C.C.B. acknowledges support from FCT through FCT contracts nr. IF/01312/2014/CP1215/CT0004. 
XB, SC, DG, MF and JL acknowledge their role as ESA-appointed CHEOPS science team members. 
This project was supported by the CNES. 
The Belgian participation to CHEOPS has been supported by the Belgian Federal Science Policy Office (BELSPO) in the framework of the PRODEX Program, and by the University of Liège through an ARC grant for Concerted Research Actions financed by the Wallonia-Brussels Federation. 
L.D. is an F.R.S.-FNRS Postdoctoral Researcher. 
This work was supported by FCT - Fundação para a Ciência e a Tecnologia through national funds and by FEDER through COMPETE2020 - Programa Operacional Competitividade e Internacionalizacão by these grants: UID/FIS/04434/2019, UIDB/04434/2020, UIDP/04434/2020, PTDC/FIS-AST/32113/2017 \& POCI-01-0145-FEDER- 032113, PTDC/FIS-AST/28953/2017 \& POCI-01-0145-FEDER-028953, PTDC/FIS-AST/28987/2017 \& POCI-01-0145-FEDER-028987, O.D.S.D. is supported in the form of work contract (DL 57/2016/CP1364/CT0004) funded by national funds through FCT. 
B.-O. D. acknowledges support from the Swiss State Secretariat for Education, Research and Innovation (SERI) under contract number MB22.00046. 
This project has received funding from the European Research Council (ERC) under the European Union’s Horizon 2020 research  grant agreement No 724427). It has also been carried out in the frame of the National Centre for Competence in Research PlanetS supported by the Swiss National Science Foundation (SNSF). DE acknowledges financial support from the Swiss National Science Foundation for project 200021\_200726.  and innovation programme (project {\sc Four Aces}. 
MF and CMP gratefully acknowledge the support of the Swedish National Space Agency (DNR 65/19, 174/18). 
DG gratefully acknowledges financial support from the CRT foundation under Grant No. 2018.2323 ``Gaseousor rocky? Unveiling the nature of small worlds''. 
M.G. is an F.R.S.-FNRS Senior Research Associate. 
MNG is the ESA CHEOPS Project Scientist and Mission Representative, and as such also responsible for the Guest Observers (GO) Programme. MNG does not relay proprietary information between the GO and Guaranteed Time Observation (GTO) Programmes, and does not decide on the definition and target selection of the GTO Programme. 
SH gratefully acknowledges CNES funding through the grant 837319. 
KGI is the ESA CHEOPS Project Scientist and is responsible for the ESA CHEOPS Guest Observers Programme. She does not participate in, or contribute to, the definition of the Guaranteed Time Programme of the CHEOPS mission through which observations described in this paper have been taken, nor to any aspect of target selection for the programme. 
This work was granted access to the HPC resources of MesoPSL financed by the Region Ile de France and the project Equip@Meso (reference ANR-10-EQPX-29-01) of the programme Investissements d'Avenir supervised by the Agence Nationale pour la Recherche. 
PM acknowledges support from STFC research grant number ST/M001040/1. 
IRI acknowledges support from the Spanish Ministry of Science and Innovation and the European Regional Development Fund through grant PGC2018-098153-B- C33, as well as the support of the Generalitat de Catalunya/CERCA programme. 
GyMSz acknowledges the support of the Hungarian National Research, Development and Innovation Office (NKFIH) grant K-125015, a a PRODEX Experiment Agreement No. 4000137122, the Lend\"ulet LP2018-7/2021 grant of the Hungarian Academy of Science and the support of the city of Szombathely. 
NAW acknowledges UKSA grant ST/R004838/1.


NCS acknowledges funding by the European Union (ERC, FIERCE, 101052347). Views and opinions expressed are however those of the author(s) only and do not necessarily reflect those of the European Union or the European Research Council. Neither the European Union nor the granting authority can be held responsible for them.

KWFL was supported by Deutsche Forschungsgemeinschaft grants RA714/14-1 within the DFG Schwerpunkt SPP 1992, Exploring the Diversity of Extrasolar Planets.

In this work we use the python package \texttt{PyDE} available at \url{https://github.com/hpparvi/PyDE}.

This research has made use of the SVO Filter Profile Service (http://svo2.cab.inta-csic.es/theory/fps/) supported from the Spanish MINECO through grant AYA2017-84089.

\end{acknowledgements}

\bibliographystyle{aa}
\bibliography{references}

\begin{appendix}

\section{Posterior distributions of the model parameters from the fit of the \tess\ transit light curves}


\begin{figure*}
    \centering
    \includegraphics[width=\linewidth]{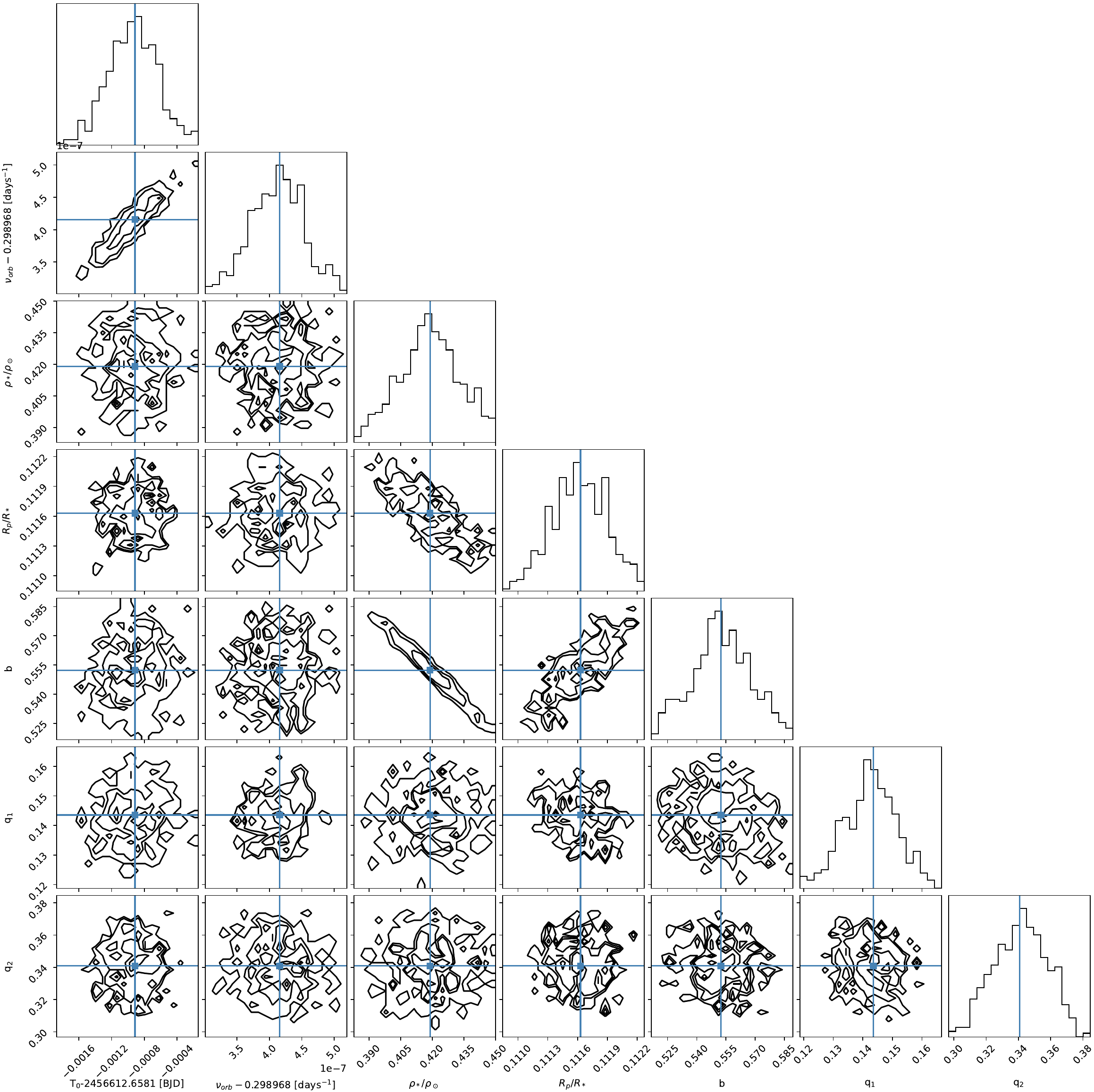}
    \caption{Corner plot of the \ac{MCMC} chains of planetary parameters from the fit of the \tess\ transits (see Sect.~\ref{sec:tessTR}). In each plot, the solid blue lines mark the MAP values.} \label{fig:tessTRcorner}
\end{figure*}

\FloatBarrier

\section{Analysis of the \tess\ transits in high cadence}\label{sec:transitsHighCadence}
In Sect.~\ref{sec:tessPer} we analyzed the \tess\ photometry at low cadence (30~min), and we found a significant difference between the transit depth between sectors 11 and 38. We speculated that the difference in transit depth is due to photospheric variability, which manifests also as a quasi-periodic signal in the LC of sector 38. This hypothesis is supported by the growing evidence that A-type stars show rotational signals due to photospheric inhomogeneities \citep[e.g., ][]{Balona2011,Bohm2015,Sikora2020}. This scenario is further supported by the fact that the \wcso\ is classified as an Am star by \citet{Hellier2019}.

\begin{figure*}
    \centering
    \includegraphics[width=.45\linewidth]{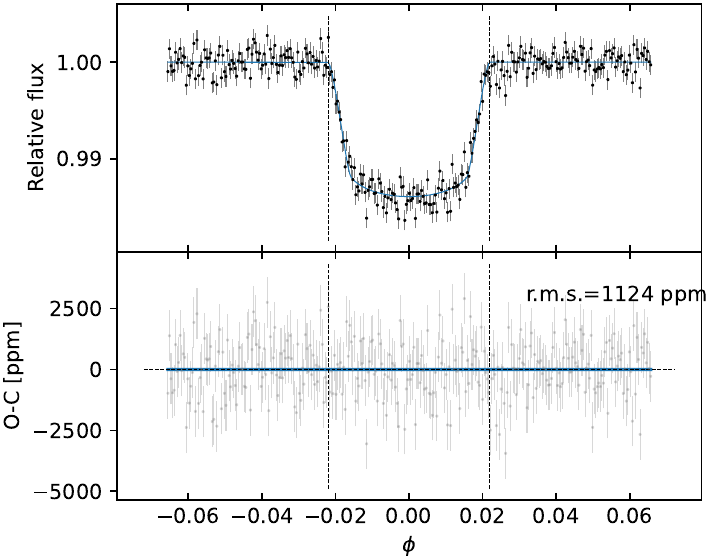}
    \includegraphics[width=.45\linewidth]{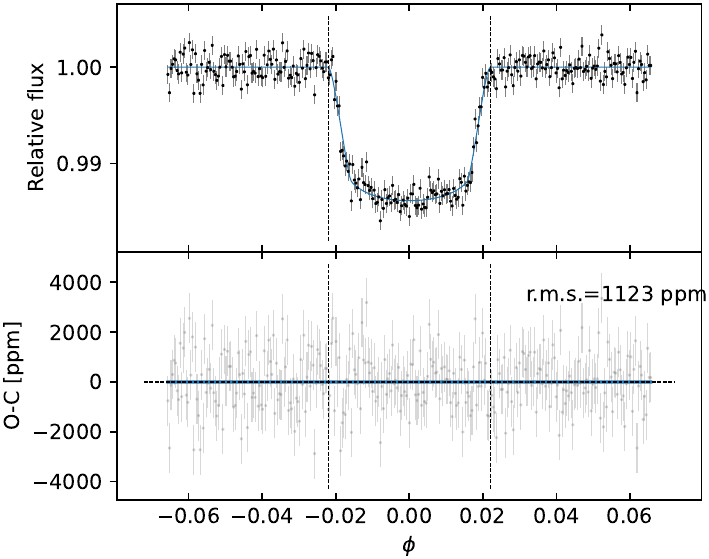}
    \caption{Detrended short cadence \ac{LC} of the first (left panel) and second (right panel) transit  observed in \tess\ sector 38. In each panel, the top plot shows the model computed with the MAP parameters in Table~\ref{tab:ephemeris} as a blue solid line. The bottom plots show the residuals of the short cadence photometry with respect to the planetary model shown in the top panel. As a guide, we plot with a blue line the smoothing of the residuals obtained with a Savitzky-Golay filter. In all plots, the vertical dashed lines mark the first and fourth contacts.}\label{fig:transit1}
\end{figure*}

\begin{figure*}
    \centering
    \includegraphics[width=.45\linewidth]{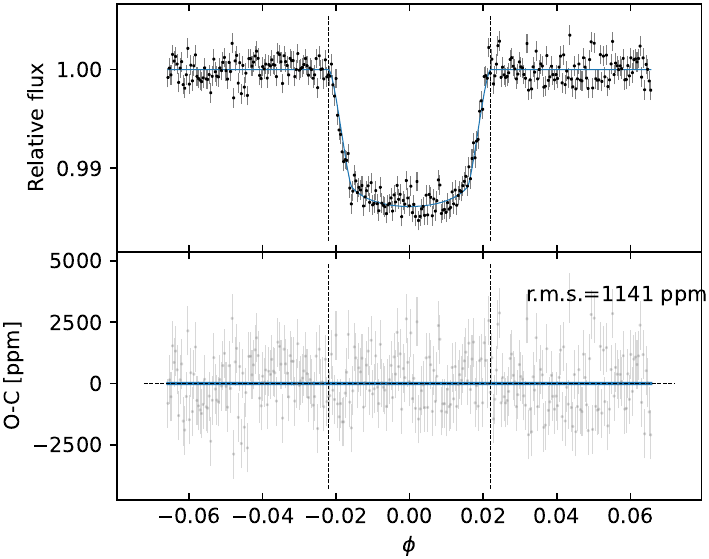}
    \includegraphics[width=.45\linewidth]{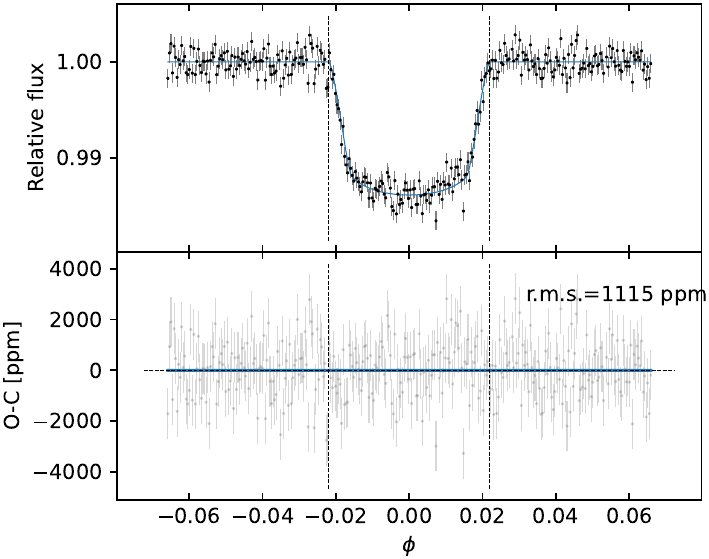}
    \caption{Detrended short cadence \ac{LC} of the third (left panel) and fourth (right panel) transit  observed in \tess\ sector 38. Details are the same as in Fig.~\ref{fig:transit1}.}\label{fig:transit3}
\end{figure*}

\begin{figure*}
    \centering
    \includegraphics[width=.45\linewidth]{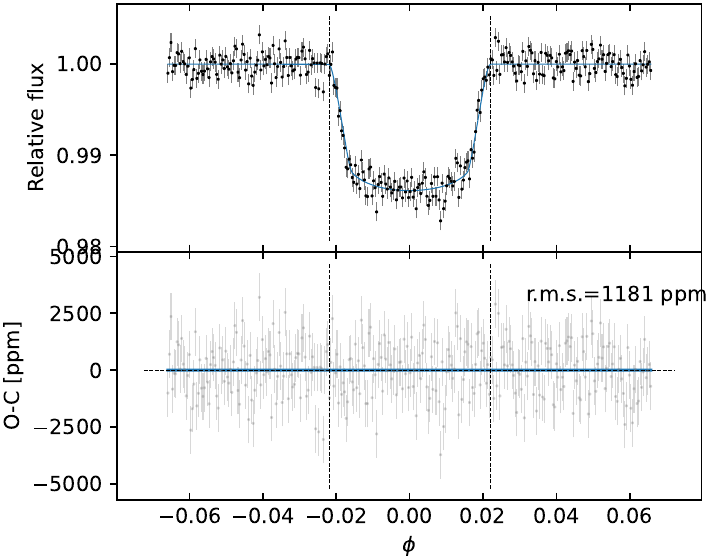}
    \includegraphics[width=.45\linewidth]{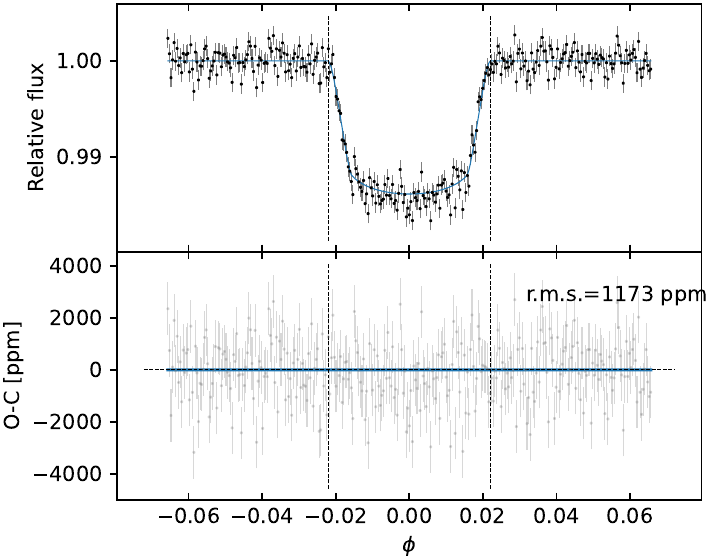}
    \caption{Detrended short cadence \ac{LC} of the fifth (left panel) and sixth (right panel) transit observed in \tess\ sector 38. Details are the same as in Fig.~\ref{fig:transit1}.}\label{fig:transit5}
\end{figure*}

\begin{figure*}
    \centering
    \includegraphics[width=.45\linewidth]{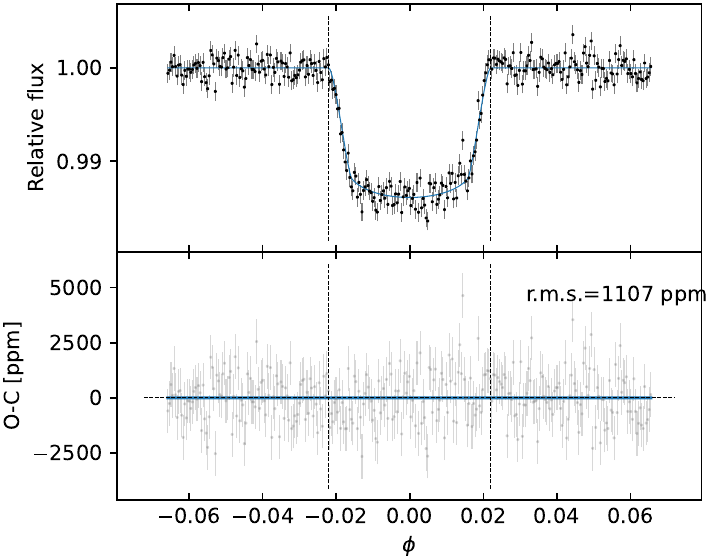}
    \includegraphics[width=.45\linewidth]{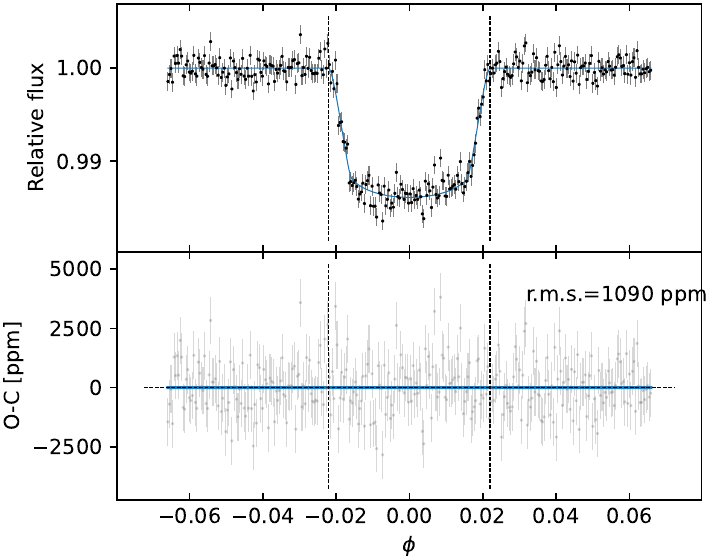}
    \caption{Detrended short cadence \ac{LC} of the seventh (left panel) and eighth (right panel) transit observed in \tess\ sector 38. Details are the same as in Fig.~\ref{fig:transit1}.}\label{fig:transit7}
\end{figure*}

To test if the visible hemisphere of the star hosts dark spots, we searched for transit anomalies in the \tess\ \ac{LC}s, that are localized bumps in the residuals of the transit fits. If present, they indicate that the planet's projection on the stellar surface crosses a darker area compared to the quiescent photosphere \citep[see for example ][]{Beky2014,Scandariato2017}. To this purpose, the low cadence photometry analyzed in Sect.~\ref{sec:tessPer} is of little help, as a finer time sampling is needed. We thus compared the high cadence (2~min) \tess\ photometry of sector 38 with the bestfit model discussed in Sect.~\ref{sec:tessPC}, computed using $R_{\rm p}/R*$=0.1141 (see Sect.~\ref{sec:tessTR}).
This comparison does not show any bump inside the transits (Figs.~\ref{fig:transit1}--\ref{fig:transit7}), and we concluded that in the eight transits observed in sector 38 there is no spot-crossing event.

We also remark that, in contrast with \citet{Rodriguez2020}, we did not detect any systematic asymmetry in the transit profile. The stellar flux distribution along the transit path is symmetric with respect to the transit center. This either indicates that the star does not show any gravity darkening, which is unlikely if the stellar rotation period of $\sim$3.2~days is confirmed, or supports the hypothesis that the star is in a pole-on geometry, which leads to a radially symmetric flux distribution over the visible stellar hemisphere.

\end{appendix}

\end{document}